\documentclass{pasa}
\title[The ISM towards HESS\,J1614$-$518 and HESS\,J1616$-$508]{A study of the interstellar medium towards the unidentified dark TeV $\gamma$-ray sources HESS\,J1614$-$518 and HESS\,J1616$-$508}
\author[J. C. Lau et al.]{J. C. Lau$^{1}$\thanks{E-mail: james.lau@adelaide.edu.au}, G. Rowell$^1$, F. Voisin$^1$, C. Braiding$^2$, M. Burton$^{2,3}$, Y. Fukui$^{4}$, S. Pointon$^5$, M. Ashley$^6$, C. Jordan$^7$ and A. Walsh$^7$\\
\affil{$^1$School of Physical Sciences, University of Adelaide, Adelaide, SA 5005, Australia}
\affil{$^2$School of Physics, University of New South Wales, Sydney, NSW 2052, Australia}
\affil{$^3$Armagh Observatory and Planetarium, College Hill, Armagh, BT61 9DG, Northern Ireland, UK}
\affil{$^4$Department of Physics, University of Nagoya, Furo-cho, Chikusa-ku, Nagoya, 464-8601, Japan}
\affil{$^5$Centre for Astrophysics and Supercomputing, Swinburne University of Technology, Hawthorn, Victoria 3122, Australia}
\affil{$^6$School of Physics, University of New South Wales, Sydney NSW 2052, Australia}
\affil{$^7$International Centre for Radio Astronomy Research, Curtin University, Bentley, WA 6845, Australia}}
\jid{PASA}
\doi{10.1017/pas.\the\year.xxx}
\jyear{\the\year}

\usepackage[authoryear]{natbib}
\bibpunct{(}{)}{;}{a}{}{,}
\usepackage{aas_macros}
\usepackage{hyperref} 
\hypersetup{colorlinks,citecolor=blue,linkcolor=blue,urlcolor=blue}

\usepackage{soul}

\begin{document}

\begin{abstract}
HESS\,J1614$-$518 and HESS\,J1616$-$508 are two tera-electron volt (TeV) $\gamma$-ray sources that are not firmly associated with any known counterparts at other wavelengths. We investigate the distribution of interstellar medium towards the TeV $\gamma$-ray sources using results from a 7\,mm-wavelength Mopra study, the Mopra Southern Galactic Plane CO Survey, the Millimetre Astronomer's Legacy Team - 45 GHz survey and [CI] data from the HEAT telescope. Data in the CO(1$-$0) transition lines reveal diffuse gas overlapping the two TeV sources at several velocities along the line of sight, while observations in the CS(1$-$0) transition line reveal several interesting dense gas features. To account for the diffuse atomic gas, archival H\textsc{i} data was taken from the Southern Galactic Plane Survey. The observations reveal gas components with masses $\sim10^3$ to $10^5$ M$_\odot$ and with densities $\sim10^2$ to $10^3$ cm$^{-3}$ overlapping the two TeV sources. Several origin scenarios potentially associated with the TeV $\gamma$-ray sources are discussed in light of the distribution of the local interstellar medium. We find no strong convincing evidence linking any counterpart with HESS\,J1614$-$518 or HESS\,J1616$-$508.

\end{abstract}

\begin{keywords}
ISM:clouds -- ISM: cosmic rays -- gamma-rays: ISM -- molecular data
\end{keywords}

\maketitle

\section{Introduction}
\label{sec:intro}
Exploration into the nature of the very high energy (VHE, $E > 100$ GeV) $\gamma$-ray sky has rapidly progressed with the use of Imaging Air Cherenkov Telescopes (IACTs). Telescopes such as the High Energy Stereoscopic System (HESS), an array of IACTs, have found many VHE $\gamma$-ray sources along the Galactic plane \citep{hess2005,hess_plane,2015ICRC...34..773D}. Many of the extended, Galactic sources have been associated with high-energy phenomena, such as pulsar wind nebulae (PWN), supernova remnants (SNRs) and binaries (\citealt{2008AIPC.1085..285R,2005A&A...442....1A,2008A&A...490..685A} etc.). However, a large population of VHE sources remain unassociated, and appear to have no clear counterparts seen in other wavelengths \citep{2015ICRC...34..773D,doi:10.1063/1.4968905}.

Astrophysical TeV $\gamma$-rays have two main mechanisms of production: the decay of neutral pions produced by the hadronic interactions between highly accelerated cosmic-ray particles and ambient interstellar medium (ISM); and the leptonic interaction of upscattering background photons via the inverse-Compton effect by high energy electrons. Understanding the distribution of the ISM towards unidentified TeV sources is thus critical in order to constrain the possible TeV $\gamma$-ray production scenarios. Here, we focus on HESS\,J1614$-$518 and HESS\,J1616$-$508, two of the most prominent unidentified TeV sources detected in the first HESS Galactic Plane Survey \citep{hess2005,hess_plane}.
\\

HESS\,J1614$-$518 is a TeV $\gamma$-ray source that was first discovered by HESS as part of a survey of the Galactic plane \citep{hess2005}. It was the brightest of the new sources discovered in the survey, with a flux 25$\%$ that of the Crab Nebula above 200 GeV. It has a TeV $\gamma$-ray spectrum that is well fit by a power-law, d$N$/dE $= N_0E^{-\Gamma}$, with a photon index $\Gamma = 2.46 \pm 0.21$. The TeV emission has an elliptical morphology, with semi-major and semi-minor axes of $14 \pm 1$ and $9 \pm 1$ arcmin respectively. It is also characterised by two peaks of emission in the Galactic North-East and Galactic South-West of the source. No immediately obvious counterpart to this source appeared in other wavelengths, and HESS\,J1614$-$518 was considered a ``dark-accelerator".

Recent preliminary results from HESS, following a systematic search for new TeV-emitting SNRs, suggest that HESS\,J1614$-$518 may have a shell-like TeV $\gamma$-ray morphology \citep{doi:10.1063/1.4968934}. As no evidence of an associated SNR has been seen so far in other wavelengths, HESS\,J1614$-$518 is currently considered a SNR candidate.

Observations towards HESS\,J1614$-$518 in X-rays made by \emph{Suzaku} revealed three X-ray sources within the TeV $\gamma$-ray source \citep{2008PASJ...60S.163M}. The first source, Suzaku Src A, is extended and is located close (within $\sim 8$ arcmin) to the brightest TeV peak of HESS\,J1614$-$518. The 2nd X-ray source, Suzaku Src B, is located towards the centre of HESS\,J1614$-$518. Additional \emph{Suzaku} observations and analysis of \emph{XMM-Newton} archival data revealed that Suzaku Src B was comprised of several point sources \citep{2011PASJ...63S.879S}. The brightest of these, XMM-Newton source B1, had the largest count rate by a factor $\sim 5$, and indicated that it was the main object of Suzaku Src B. \cite{2011PASJ...63S.879S} postulated that HESS\,J1614$-$518 could be an SNR associated with an Anomalous X-ray Pulsar (AXP). This scenario has XMM-Newton source B1 as the AXP produced by a supernova explosion, and Suzaku source A as the shocked region of the SNR. The other X-ray source found by \emph{Suzaku}, Suzaku source C, was found to be a late-type B star \citep{2008PASJ...60S.163M}.

The X-ray telescope (XRT) aboard \emph{Swift} observed the region towards HESS\,J1614$-$518 and found six point-like X-ray sources \citep{2007ATel.1047....1L}. Four of these (Swift sources 1, 2, 3 and 5 in \citealt{2007ATel.1047....1L})  were identified as stars, while the others (Swift sources 4 and 6) remain unidentified. The Swift source 1 and 4 are coincident with Suzaku source B and C respectively. Suzaku source A was not seen by the \emph{Swift} XRT in these observations.

A GeV source is seen towards HESS\,J1614$-$518 by the \emph{Fermi} Large Area Telescope (\emph{Fermi}-LAT). Designated 3FGL\,J1615.3$-$5146e in the 3rd Fermi point source catalogue \citep{2015ApJS..218...23A}, the extended source was classified as being `disk-like', and has a relative large diameter of $\sim 0.8^{\circ}$ which covers a major fraction of HESS\,J1614$-$518.

\begin{figure}[!ht]
\centering
\includegraphics[width=\linewidth]{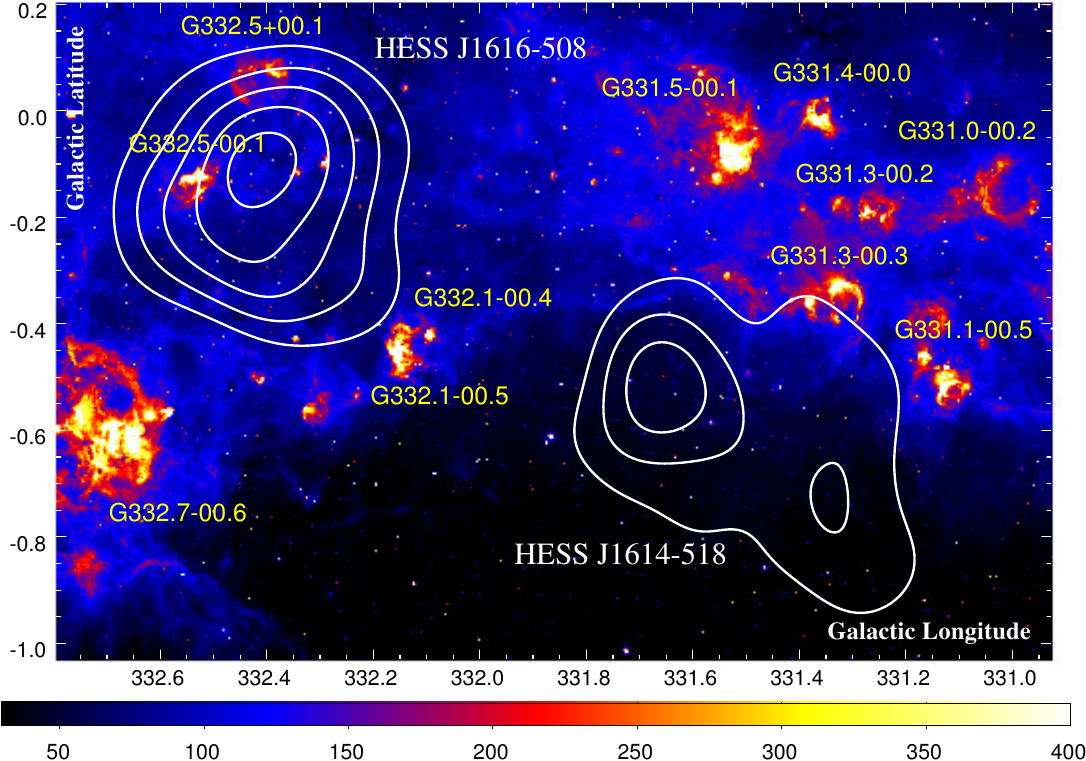}
\caption[what]{\textit{Spitzer} GLIMPSE 8.0 $\mu$m image [MJy\,sr$^{-1}$]  towards HESS\,J1614$-$518 and HESS\,J1616$-$508 \citep{2009PASP..121..213C}. White contours are HESS excess counts contours at the 30, 45, 60, 75 and 90 levels \citep{hess_plane}. Nearby H\textsc{ii} regions are labelled in yellow \citep{2003A&A...397..213P}.}
  \label{fig:spitzer_hi}
\end{figure}

Several H\textsc{ii} regions and molecular cloud complexes appear to the Galactic-North of HESS\,J1614$-$518. Figure \ref{fig:spitzer_hi} is a \textit{Spitzer} GLIMPSE 8.0 $\mu$m image \citep{2009PASP..121..213C} of the region towards HESS\,J1614$-$518 and the neighbouring TeV source HESS\,J1616$-$508, with nearby H\textsc{ii} regions labelled in yellow. While there are several H\textsc{ii} regions in the vicinity of HESS\,J1614$-$518, none appear to overlap the TeV source.

A multi-wavelength counterpart study of HESS\,J1614$-$518 was conducted by \cite{2008AIPC.1085..241R}. The known pulsars towards HESS\,J1614$-$518 were thought likely not responsible for the TeV $\gamma$-ray emission due to their insufficient spin-down powers, though a small fractional contribution could not be ruled out. A possible association between HESS\,J1614$-$518 and the young open stellar cluster Pismis 22 \citep{Piatti} was suggested in scenarios where stellar winds from several B-type stars or undetected SNRs from deceased members of the cluster would accelerate cosmic-rays that would interact with ambient gas to produce $\gamma$-rays hadronically. 

\cite{2011ApJ...740...78M} used the CANGAROO-III telescopes to study the TeV $\gamma$-ray emission towards HESS\,J1614$-$518, and also investigated the plausibility of several radiation mechanisms. A leptonic scenario based on an undetected SNR was rejected as it was not able to reproduce the observed spectral energy distribution (SED) in $\gamma$-rays. On the other hand, hadronic models that involved either a SNR or stellar winds from Pismis 22 were found to produce a good reproduction of the SED. Certain requirements, however, on the number density of the ISM were needed, and the initial investigation of the Nanten $^{12}$CO(1$-$0) survey data by \cite{2008AIPC.1085..241R} in this region had revealed no obvious overlapping molecular clouds along the line of sight up to a kinematic distance of $\sim 6$\,kpc. More detailed and higher resolution ISM data were needed in order to test the validity of these models.
\\

\textit{HESS\,J1616$-$508} is located less than a degree away from HESS\,J1614$-$518. It too was discovered as part of the HESS Galactic plane survey \citep{hess2005}, with a $\gamma$-ray flux 19$\%$ that of the Crab Nebula above 200 GeV. The TeV spectrum is fit by a power-law model with a photon index $\Gamma = 2.35\pm0.06$. It has a roughly circular extended morphology, with an angular diameter of $\sim16$ arcmin. There are two SNRs near HESS\,J1616$-$508, Kes 32 (G332.4+0.1) and RCW 103 (G332.4$-$0.4), located 17 and 13 arcmin away respectively. Due to the distance from the centre of HESS\,J1616$-$508, an association between the TeV source and either SNR has been disfavoured \citep{Landi2007,Kargaltsev2008}.

Three pulsars are seen towards the vicinity of HESS\,J1616$-$508. Two of them, PSR J1616$-$5109 and PSR J1614$-$5048 are unlikely to be associated with the TeV emission due to their large separation from HESS\,J1616$-$508. On the other hand, the young ($\sim 8$ kyr) pulsar PSR\,J1617$-$5055 \citep{Kaspi1998} and associated pulsar wind nebula (PWN) have been suggested as candidate objects responsible for the TeV emission \citep{Landi2007,Aharonian2008,2011ICRC....6..202T,2013ApJ...773...77A}. PSR J1617$-$5055 has a spin-down power sufficient to supply the appropriate energetics, but is offset from the centre of the TeV source by $\sim 9$ arcmin. Observations in X-rays by \emph{Chandra} revealed a faint PWN extending from PSR J1617$-$5055 \citep{Kargaltsev2008}. However, PWNs associated with offset TeV $\gamma$-ray emission typically have an extension of the X-ray emission towards said $\gamma$-rays, and the \emph{Chandra} observations found no evidence of any X-ray asymmetry in the PWN towards HESS\,J1616$-$508. Recent analysis of three \emph{Chandra} observations covering most of HESS\,J1616$-$508 was performed by \cite{2017ApJ...841...81H}, finding 56 X-rays sources within the fields. Many of the sources were identified as active galactic nuclei (AGN) and non-degenerate stars, but none were found to be promising counterparts to the TeV source.

Observations by \emph{Fermi}-LAT reveal a GeV $\gamma-$ray source, 3FGL\,J1616.2$-$5054e towards HESS\,J1616$-$508 \citep{2015ApJS..218...23A}. It has a diameter of $\sim0.6^{\circ}$ and is positionally coincident with the TeV source.

A number of  H\textsc{ii} regions are seen towards HESS\,J1616$-$508, with several overlapping the TeV source as shown in Figure\,\ref{fig:spitzer_hi}.
\\

To better understand the origins of HESS\,J1614$-$518 and HESS\,J1616$-$508, a detailed understanding of the distribution and characteristics of the ISM towards these two sources is required. In order to achieve this, we have used molecular line data taken by the Mopra radio telescope and the Australia Telescope Compact Array (ATCA). The diffuse ($\overline{n}\gtrsim10^3$ cm$^{-3}$) gas towards the two TeV sources were traced as part of the Mopra Southern Galactic Plane CO Survey \citep{2013PASA...30...44B}. Data were taken from the \mbox{Millimetre Astronomer's Legacy Team - 45 GHz} (MALT-45) survey \citep{Jordan2015} which targeted the dense ($\overline{n}\gtrsim10^4$ cm$^{-3}$) gas tracers in the 7\,mm wavelength band including CS(1$-$0) and SiO(1$-$0, v=0). As the MALT-45 survey did not extend to encompass HESS\,J1614$-$518 entirely, we took further observations in the 7\,mm wavelength band with Mopra to complete the coverage in the dense gas tracers.

Section 2 describes the parameters and reduction process involved with the data taken with the Mopra radio telescope, as well as the parameters of the data taken from the MALT-45 survey. Section 3 describes the gas parameter calculations that we apply to the data. In Section 4, we present our findings of the the distribution of the ISM towards the TeV sources. Finally, in Section 5, we discuss our results and the implications they have on the possible production scenarios for HESS\,J1614$-$518 and HESS\,J1616$-$508.

\section{Datasets, observations and data reduction}
High resolution data in the 7\,mm wavelength band was taken from the MALT-45 survey \citep{Jordan2015}. This survey made use of ATCA to survey an area of 5 square-degrees along the Galactic plane ($l = 330^{\circ} - 335^{\circ}$, $b=\pm0.5$). Across the 7\,mm band, the survey FWHM ranged from 57$''$ (49 GHz) to 66$''$ (43 GHz) with a velocity resolution of $\sim0.2$ km\,s$^{-1}$. Complete details of the survey are presented in the aforementioned paper. The survey area of MALT-45 completely covered HESS\,J1616$-$508, but only covered the northern-half of HESS\,J1614$-$518.

A 7\,mm targeted study was carried out with the Mopra radio telescope to complement the MALT-45 survey, completing the coverage of HESS\,J1614$-$518. The Mopra observations were taken between September and November 2013. Two Mopra `On-the-fly' (OTF) maps were taken, each with a size of 20$'$ by 20$'$. Together, this formed a 40$'$ by 20$'$ map which was centred on [$l,b$]=[331$^{\circ}$.50, $-0^{\circ}.67$]. For these observations, we used the same scan settings as per \cite{2017MNRAS.464.3757L}.

These observations utilised the Mopra spectrometer, MOPS, in its `zoom' mode, allowing for recording of sixteen sub-bands simultaneous, each with 4096-channels and a 137.5 MHz bandwidth. The beam FWHM of Mopra in the 7\,mm band is $\sim 1'$ at 49 GHz, with a velocity resolution of $\sim0.2$ km\,s$^{-1}$. The specific molecular line transitions that were targeted by MOPS are listed along with the achieved $T_{\text{RMS}}$ levels in Table\,\,\ref{table:mops}. We note that MALT-45 had increased sensitivities compared with that of our Mopra observations. In particular, the MALT-45 $T_{\text{RMS}}$ for the CS(1$-$0) line was $\sim 0.034$ K compared with the $\sim 0.09$ K we achieved with Mopra.

\begin{table}
\caption{The set-up for the Mopra Spectrometer (MOPS) for the 7\,mm observations. The targeted molecular lines, targeted frequencies and achieved mapping $T_{\text{RMS}}$ are displayed.}
\begin{center}
\label{table:mops}
\begin{tabular}{cccc}
\hline
 Molecular line & Frequency & $T_{\text{RMS}}$\\
  & (GHz) & (K/channel)\\
\hline 
 $^{30}$SiO(J=1-0, v=0) & 42.373365 & 0.07\\ 
 SiO(J=1-0, v=3) & 42.519373 & 0.07\\ 
 SiO(J=1-0, v=2) & 42.820582 & 0.07\\ 
 $^{29}$SiO(J=1-0, v=0) & 42.879922 & 0.07\\ 
 SiO(J=1-0, v=1) & 43.122079 & 0.07\\ 
 SiO(J=1-0, v=0) & 43.423864 & 0.07\\ 
 CH$_{3}$OH-I & 44.069476 & 0.07\\ 
 HC$_{7}$N(J=40-39) & 45.119064 & 0.07\\ 
 HC$_{5}$N(J=17-16) & 45.264750 & 0.07\\ 
 HC$_{3}$N(J=5-4, F=4-3) & 45.490264 & 0.08\\ 
 $^{13}$CS(J=1-0) & 46.247580 & 0.08\\ 
 HC$_{5}$N(J=16-15) & 47.927275 & 0.08\\ 
 C$^{34}$S(J=1-0) & 48.206946 & 0.09\\ 
 OCS(J=4-3) & 48.651604 & 0.09\\
 CS(J=1-0) & 48.990957 & 0.09\\ 
\hline
\end{tabular}
\end{center}
\end{table}

Data in the CO(1$-$0) lines was provided by the Mopra Southern Galactic Plane CO Survey \citep{2013PASA...30...44B, 2015PASA...32...20B}. The survey targets the $^{12}$CO, $^{13}$CO and C$^{18}$O $J$ = 1$-$0 molecular lines within the fourth quadrant of the Galaxy ($l=305^{\circ}$ to $345^{\circ}$, and $b=\pm0^{\circ}.5$). The beamsize of this survey is 0$'$.6 , with a velocity resolution of 0.1 km\,s$^{-1}$. The $T_\text{RMS}$ for the $^{12}$CO and $^{13}$CO lines is $\sim1.5$\,K $\sim0.7$\,K respectively. We refer to the aforementioned papers for further details about the survey. 

We used ATNF analysis software, \textsf{Livedata}\footnote{http://www.atnf.csiro.au/computing/software/livedata/ \label{foot:1}}, \textsf{Gridzilla}$^{\ref{foot:1}}$, and \textsf{Miriad}\footnote{http://www.atnf.csiro.au/computing/software/miriad/}, together with custom \textsf{IDL} routines, in order to reduce and perform analysis on the OTF mapping data. Using \textsf{Livedata}, we calibrated the spectra with the reference OFF position, and then subtracted the baseline using a polynomial fit. Combining the data from separate scans using \textsf{Gridzilla}, we created three-dimensional cubes for each sub-band. Finally, the integrated emission maps were generated from the cubes via the use of \textsf{Miriad} and custom \textsf{IDL} routines.

\section{Spectral line analysis}
\label{sec:line_analysis}
The spectral line analyses performed on the CO(1$-$0), CS(1$-$0) and H\textsc{i} data to calculate gas mass and density parameters are outlined in \cite{2017MNRAS.464.3757L}, and are summarised here for completeness.

Spectral components and features were fit with Gaussian functions, and the integrated intensity of the line emission was then used to calculate the average column density of molecular hydrogen, $\overline{N_{\text{H}_{2}}}$. From here, the mass of the gas in the region of interest is estimated by $M = \mu m_{\text{H}} \overline{N_{\text{H}_{2}}} A$, where $m_{\text{H}}$ is the mass of a single hydrogen atom and $A$ is the cross-sectional area of the region in which the spectra was extracted from. The average molecular weight $\mu$ is taken to be 2.8 to account for the assumed $\sim 20\%$ helium content. Within the region of interest, the average number density, $\overline{n}$, is estimated by assuming a geometry with depth equal to the average height and width.

\subsection{CO}
\label{sec:co_line_analysis}
We convert the $^{12}$CO(1$-$0) integrated brightness temperature to an average H$_{2}$ column density in a region via the relation $\overline{N_{H_{2}}} = X_{^{12}\text{CO(1-0)}} W_{^{12}\text{CO(1-0)}}$, where $W_{^{12}\text{CO(1-0)}}$ is the integrated $^{12}$CO(1$-$0) intensity and $X_{^{12}\text{CO(1-0)}}$ is the $^{12}$CO(1$-$0) X-factor. In this work, we adopt the $^{12}$CO(1$-$0) X-factor $X_{^{12}\text{CO(1-0)}} \sim 1.5 \times 10^{20}$ cm$^{-2}$(K\,km/s)$^{-1}$ \citep{Strong2004}. For simplicity, we apply the same method to convert $^{13}$CO(1$-$0) integrated brightness temperatures to average H$_{2}$ column densities, using the $^{13}$CO(1$-$0) X-factor $X_{^{13}\text{CO(1-0)}} \sim 4.9 \times 10^{20}$ cm$^{-2}$(K\,km/s)$^{-1}$ \citep{Simon2001}.
Following \cite{2013PASA...30...44B}, we calculate the optical thickness of the $^{12}$CO line, $\tau_{12}$, by comparing the $^{12}$CO and $^{13}$CO lines. In the limit where the $^{12}$CO and $^{13}$CO lines are optically thick and optically thin respectively, $\tau_{12}$ can be given by:
\begin{equation}
\tau_{12} = \dfrac{X_{12/13}}{R_{12/13}}
\end{equation}
where $R_{12/13}$ is the ratio between the brightness temperatures of the $^{12}$CO and $^{13}$CO emission, with $X_{12/13}$ = [$^{12}$C/$^{13}$C] being the isotope abundance ratio. Using results presented in \cite{1982A&A...109..344H}, the abundance ratio was taken to be $X_{12/13} = 5.5R + 24.2$, where $R$ is the galactocentric radius given in kpc. In the case where no $^{13}$CO was detected, we take an upper limit on the $^{13}$CO peak intensity to be the RMS sensitivity of the data ($\sim0.7$\,K, \citealt{2013PASA...30...44B}).

\subsection{CS}
\label{sec:cs_line_analysis}
Transitions in the 7\,mm CS(1$-$0) line gave a complementary probe of the denser gas ($\overline{n}\gtrsim10^4$ cm$^{-3}$, \citealt{1999ARA&A..37..311E}) in regions of interest. The CS(1$-$0) optical depth was found using the ratio between the CS(1$-$0) and C$^{34}$S(1$-$0) lines in regions where detections were made in C$^{34}$S(1$-$0). We adopted the \mbox{[CS]/[C$^{34}$S]} ratio of 22.5, and calculated the optical depth via Equation 1 of \cite{1994A&A...288..601Z}. Where no C$^{34}$S(1$-$0) was detected, the CS(1$-$0) was assumed to be optically thin.

Using Equation 9 from \cite{1999ApJ...517..209G}, together with the optical depth and the integrated line intensity, we calculate the column density of CS(J=1). We assume Local Thermodynamic Equilibrium (LTE) at a rotational temperature of $T_{\text{rot}} = 10$ K, typical of cold and dense molecular clouds, to convert the column density of CS(J=1) to total CS column density, $N_{\text{CS}}$. Here, $N_{\text{CS}}$ is $\sim 3.5$ times the CS(J=1) column density. A small systematic error in $N_{\text{CS}}$ is produced by this temperature assumption ($\sim 20\%$ for a $50\%$ change in T$_{\text{rot}}$).

We note that the abundance ratio between molecular hydrogen and CS molecules in dense molecular clumps can vary by an order of magnitude between $10^{-9}$ and $10^{-8}$ \citep{1987ASSL..134..561I}. In this work, we adopt the CS to H$_{2}$ abundance ratio $X_{\text{CS}} \sim 1 \times 10^{-9}$ \citep{Frerking1980} which is typical of dense quiescent gas. As such the calculated gas parameters presented here should be considered as upper limits.

\subsection{HI}
H\textsc{i} data towards HESS\,J1614$-$518 and HESS\,J1616$-$508 was obtained from the Southern Galactic Plane Survey (SGPS) \citep{SGPS}. The column density of atomic H\textsc{i}, $N_{\text{H}\textsc{i}}$, was calculated via the relation $\overline{N_{\text{H}\textsc{i}}} = X_{\text{H}\textsc{i}} W_{\text{H}\textsc{i}}$, where $W_{\text{H}\textsc{i}}$ is the integrated H\textsc{i} intensity and the conversion factor $X_{\text{H}\textsc{i}} = 1.823 \times 10^{18}$ cm$^{-2}$ (K\,kms$^{-1}$)$^{-1}$ \citep{Dickey1990}. Combining this with the molecular hydrogen column density, $N_{\text{H}_2}$, we are able to estimate the total hydrogen column density as $N_{\text{H}} = N_{\text{H}\textsc{i}} + 2N_{\text{H}_{2}}$.

\section{Results}
\label{sec:results}
The distribution of the ISM towards HESS\,J1614$-$518 and HESS\,J1616$-$508 is presented in this section. We consider the morphology of the gas towards each TeV source separately. $^{12}$CO(1$-$0) and $^{13}$CO(1$-$0) line emission data was taken from the Mopra Galactic Plane Survey and was used to study the diffuse molecular hydrogen gas distribution. 7\,mm wavelength data from MALT-45 and our targeted 7\,mm Mopra observations were used to reveal the denser gas as traced by detections in the CS(1$-$0) and C$^{34}$S(1$-$0) lines. Detections in the thermal SiO(J=1$-$0, v=0) line, which is often excited behind shocks that move through molecular clouds (e.g. \citealt{1992A&A...254..315M,1996MNRAS.280..447F}), were also found in the dataset.

To estimate the distance to the ISM traced by the line emission, we use the Galactic rotation model from \cite{Brand1993} to obtain the kinematic distance, based on the detection velocity along the line-of-sight ($v_{\text{LSR}}$). In the absence of firm evidence to resolve the near/far distance ambiguities, we have assumed the near solution in our calculations as an approximation, as gas closer to us is more likely to be seen.

\subsection{ISM towards HESS\,J1614$-$518}
\label{sec:result_1614}
\begin{figure}[!ht]
\centering
\includegraphics[width=\linewidth]{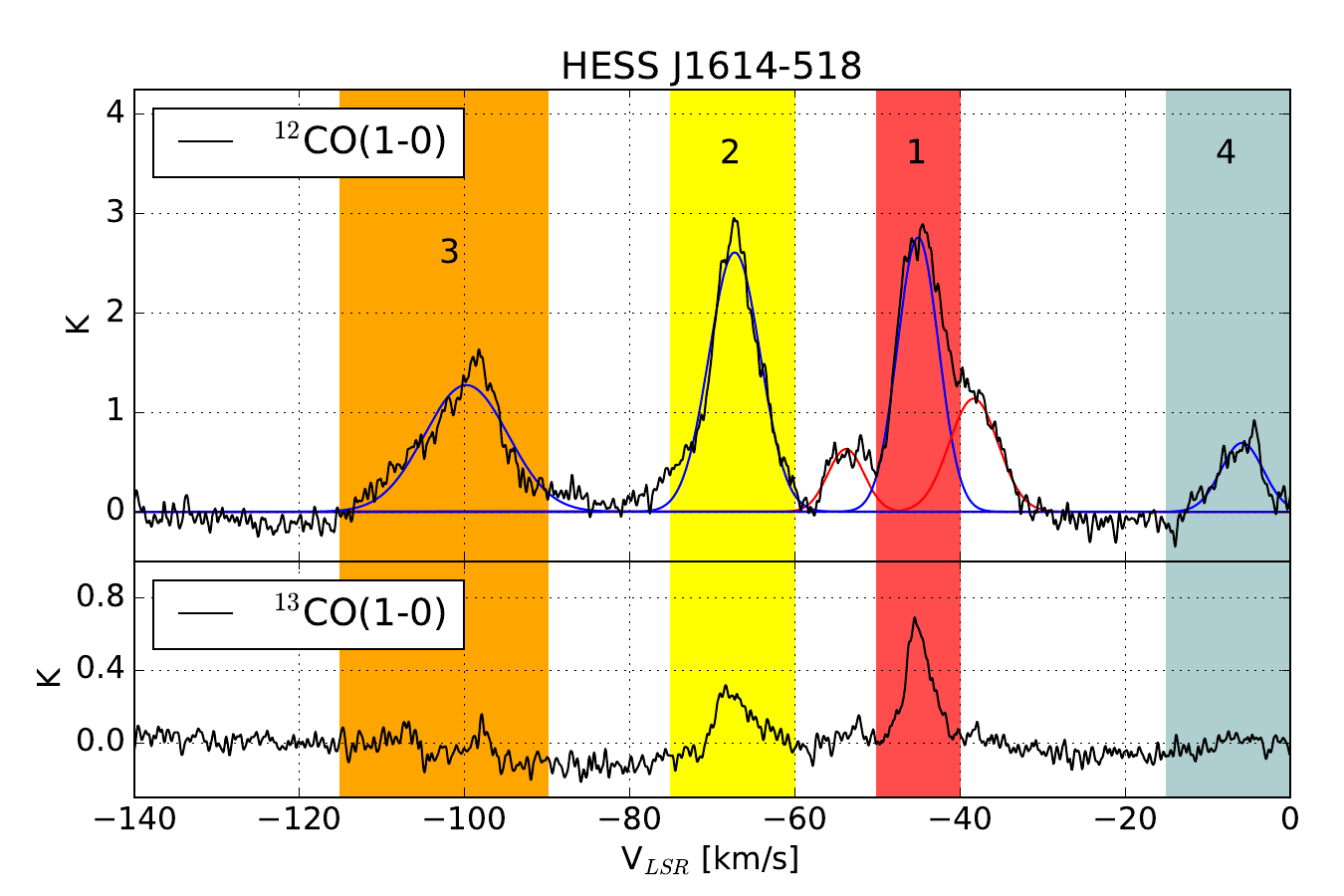}
\caption[what]{The average spectra of $^{12}$CO(1$-$0) (top) and $^{13}$CO(1$-$0) (bottom) emission within the RMS extent of HESS\,J1614$-$518 as described in \cite{hess_plane}. The velocity intervals of the components used in the integrated images shown in Figure \ref{fig:1614_co} are indicated by the shaded rectangles. Overlaid blue and red lines are the Gaussian fits to the emission, as described in text.}
  \label{fig:1614_co_spectra}
\end{figure}

\begin{figure*}[!htb]
\centering
\includegraphics[width=0.75\linewidth]{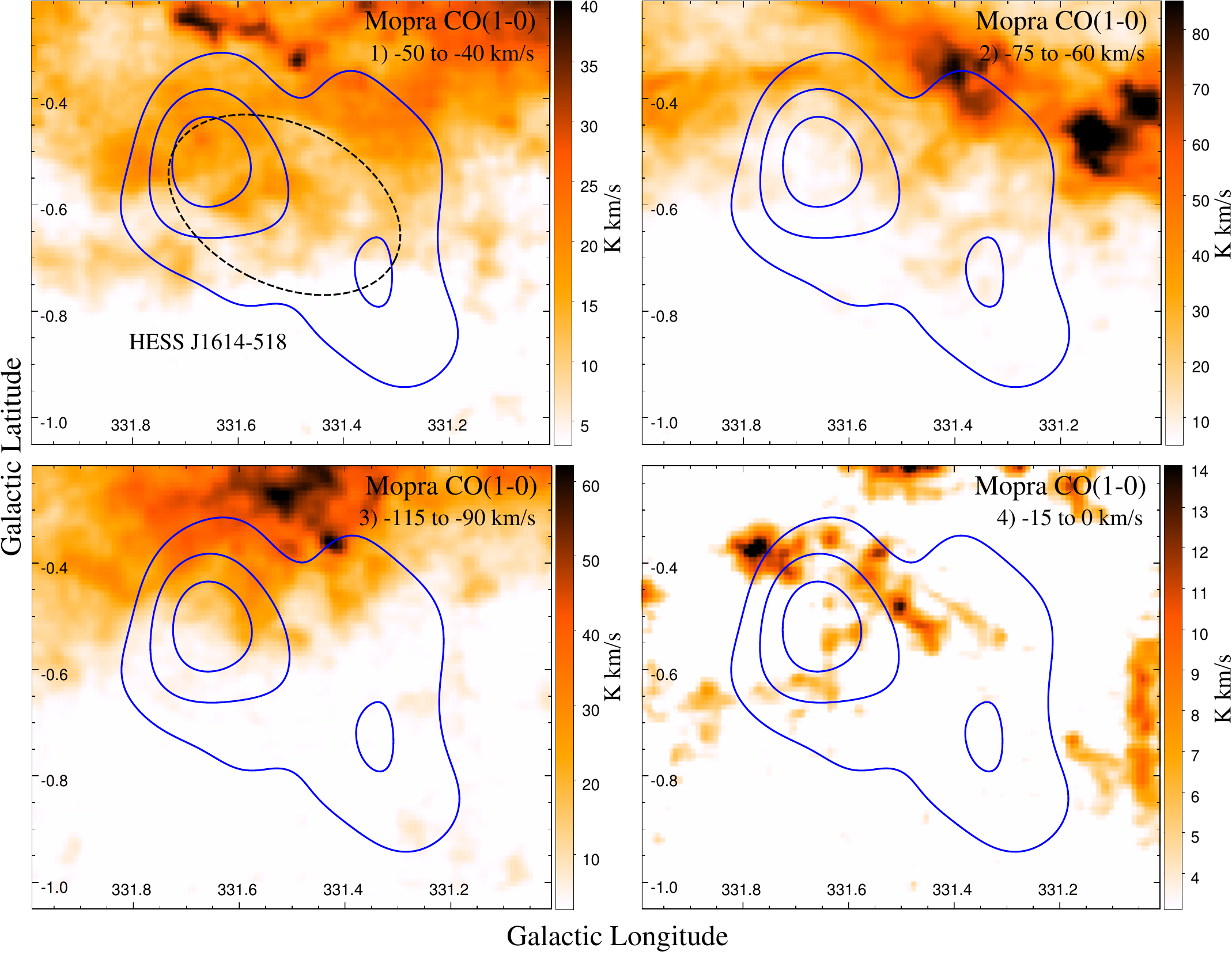}

\vspace{0.5cm}
\includegraphics[width=0.75\linewidth]{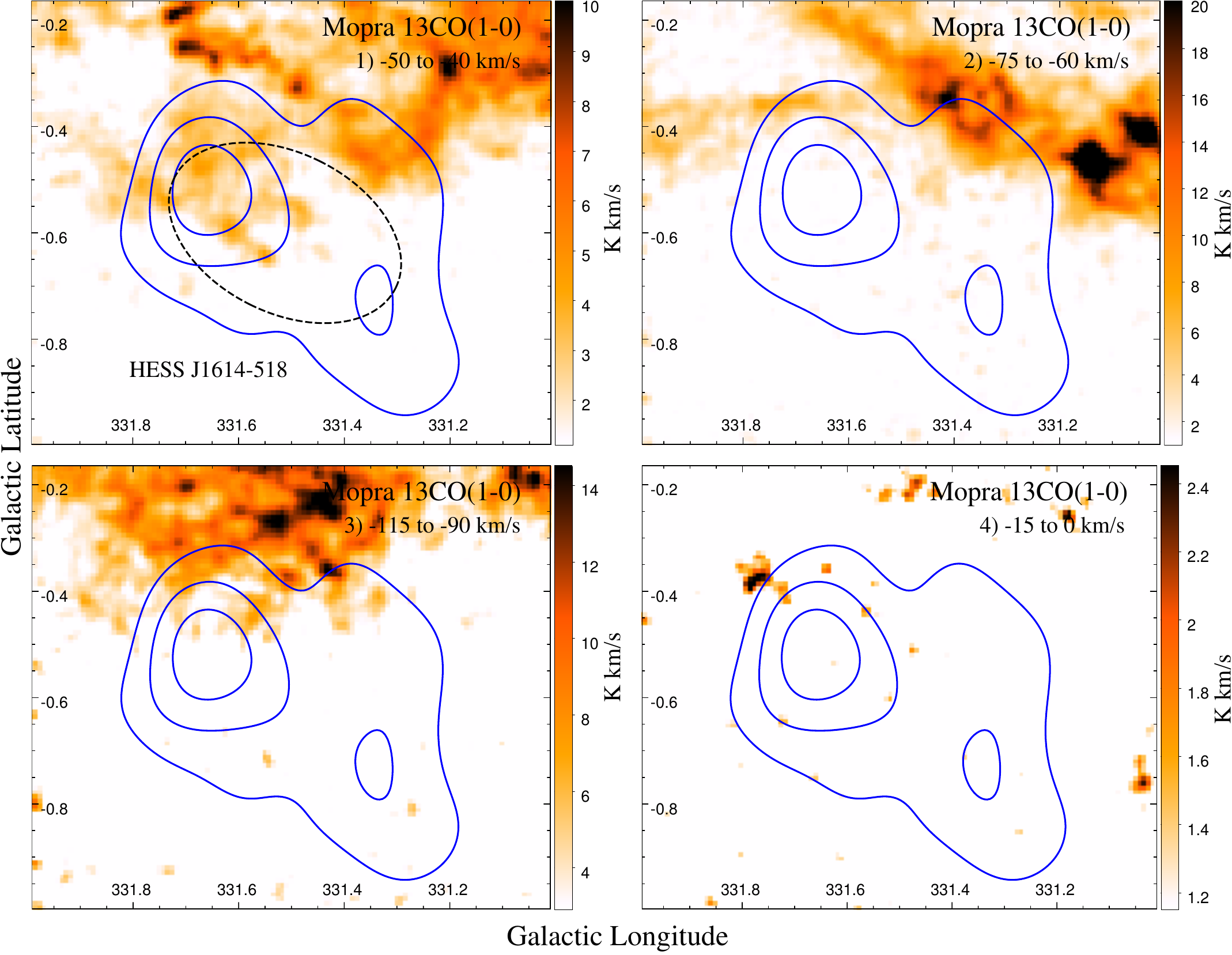}
\caption[what]{Mosaic of $^{12}$CO(1$-$0) and $^{13}$CO(1$-$0) integrated intensity images [K\,km\,s$^{-1}$] within the labelled velocity intervals. Overlaid are HESS excess counts contours (blue) towards HESS\,J1614$-$518 at the 30, 45 and 60 levels. The dashed black ellipse in the top left panel of both mosaics is the elliptical extent of HESS\,J1614$-$518 as described in \cite{hess_plane}.  The average CO(1$-$0) spectra within this region is displayed in Figure \ref{fig:1614_co_spectra}.}
  \label{fig:1614_co}
\end{figure*}

CO(1$-$0) line emission is seen overlapping HESS\,J1614$-$518 in several kinematic velocity intervals along the line-of-sight. Figure \ref{fig:1614_co_spectra} displays the average spectra of the $^{12}$CO(1$-$0) and $^{13}$CO(1$-$0) emission within the reported root-mean-squared (RMS) extent of HESS\,J1614$-$518 as described in \cite{hess_plane}, which is indicated by the dashed ellipses in Figure \ref{fig:1614_co}. Overall, the spectra indicates that there are three main velocity ranges in which emission is prominent; $v_{\text{LSR}} \sim -50$ to $-40$ km\,s$^{-1}$, $v_{\text{LSR}} \sim -75$ to $-60$ km\,s$^{-1}$ and $v_{\text{LSR}} \sim -115$ to $-90$ km\,s$^{-1}$, which we have denoted as components 1, 2, and 3 respectively. Additionally, one minor component of emission is seen in the $v_{\text{LSR}} \sim -15$ to $0$ km\,s$^{-1}$ range, which we denote as component 4. For component 1, there appears to be several blended features in the spectra, and we have chosen the velocity range that encompasses the dominant feature of interest. For the other components, we have chosen velocity ranges that cover all the emission, as the nature and degree of blending is more difficult to discern.

The $^{12}$CO(1$-$0) spectrum was fit by a series of Gaussian functions to calculate the mass and density parameters of the molecular gas following \S\ref{sec:line_analysis}. Component\,\,1 was fit with three Gaussian functions, as while the dominant feature was blended with features on both the positive and negative velocity sides, each feature is clearly resolvable. We use the Gaussian corresponding to the dominant feature in our calculations to minimise the contamination from the other blended features. For the other components, we cannot resolve the individual features that may be present, and it is unclear to the degree at which blending is occurring. For this reason, we approximate each spectrum using a single Gaussian function. The derived masses and densities from these fits and those obtained by simply integrating the raw spectrum over the velocity ranges differ by less than 10$\%$ in all cases. The Gaussian functions used have been overlaid on the $^{12}$CO(1$-$0) spectrum in Figure \ref{fig:1614_co_spectra}. Blue functions indicate those that were used in the mass and density calculations for the corresponding labelled components, while red functions are those that were fit to the extra blended features. The parameters of the blue functions as well as the calculated gas parameters are displayed in Table\,\,\ref{table:1614_ave_param}.

Figure \ref{fig:1614_co} shows a mosaic of integrated $^{12}$CO(1$-$0) and $^{12}$CO(1$-$0) emission images towards HESS\,J1614$-$518 in the velocity ranges of each component.

In component 1 ($v_{\text{LSR}}$ = $-50$ to $-40$ km\,s$^{-1}$), emission in $^{12}$CO(1$-$0) is seen overlapping most of HESS\,J1614$-$518. This emission appears not to be a localized molecular cloud as it extends North beyond the TeV source as part of widespread $^{12}$CO(1$-$0) emission. The $^{13}$CO(1$-$0) emission in this component is seen mainly towards the TeV $\gamma$-ray peak, and also extends North beyond the TeV source.

There is some overlap between $^{12}$CO(1$-$0) emission and HESS\,J1614$-$518 in component 2 ($v_{\text{LSR}}$ = $-75$ to $-60$ km\,s$^{-1}$), mainly in the Galactic North and North-west of the TeV source. However, there appears to be very little overlap in the $^{13}$CO(1$-$0) emission. Two regions of more intense emission are seen clearly in the $^{12}$CO(1$-$0) and $^{13}$CO(1$-$0), appearing at the Galactic North-West edges of HESS\,J1614$-$518. That being said, neither of the two features are likely to be associated with the HESS source, as they have no morphological correspondence with $\gamma$-ray emission. These features are coincident, and morphologically similar, to the H\textsc{ii} regions G331.3-00.3 and G331.1-00.5 which are labelled in Figure \ref{fig:spitzer_hi}. It is likely that the CO(1$-$0) emission here traces gas associated with those H\textsc{ii} regions.

The emission in $^{12}$CO(1$-$0) and $^{13}$CO(1$-$0) appearing in component 3 ($v_{\text{LSR}}$ = $-115$ to $-90$ km\,s$^{-1}$) only overlap a small portion of the TeV source in the Galactic North and North-East regions.

Component 4 ($v_{\text{LSR}}$ = $-15$ to $0$ km\,s$^{-1}$) has the weakest CO(1$-$0) emission feature detected towards HESS\,J1614$-$518. Some scattered gas is seen in $^{12}$CO(1$-$0) which overlaps the Galactic North-East region of the TeV source, while almost no emission is seen in the $^{13}$CO(1$-$0).

\begin{table*}
  \centering
  \caption{The parameters of the $^{12}$CO(1$-$0) line emission and the calculated physical parameters for the gas within the indicated aperture in Figure \ref{fig:1614_co} towards HESS\,J1614$-$518. Gaussian fits to the spectra were used to determine the line-of-sight velocity, $v_{\text{LSR}}$, line-width (full-width-half-maximum), $\Delta v_\text{FWHM}$, and peak intensity, $T_{\text{peak}}$. The $^{12}$CO/$^{13}$CO abundance ratio, $X_{12/13}$, and optical depth were found following \S\ref{sec:co_line_analysis}. The assumed distances, $d_0$, used in mass and density calculations are the near solutions derived from the Galactic rotation curve presented in \cite{Brand1993}. Calculated mass and density values can be scaled for an arbitrary distance, $d$, using a factor of $(d/d_0)^2$ and $(d/d_0)^{-1}$ respectively.}
  \begin{tabular}{cccccccccc}
  \hline 
  
Component & Distance & $v_{\text{LSR}}$ & $\Delta v_\text{FWHM}$ & $T_{\text{peak}}$ & $X_{12/13}$ & Optical & $\overline{N_{H_{2}}}$ & Mass & $\overline{n}$ \\ 
 & (kpc) & (km/s) & (km/s) & (K) & & depth & ($10^{21}$ cm$^{-2}$) & ($10^4$ M$_{\odot}$) & ($10^{2}$ cm$^{-3}$) \\
  \hline 
  
1 & 3.1 & $-45.1$ & 5.9 & 2.8 & 57.0 & 12.6 & 2.6 & 1.9 & 1.8 \\
2 & 4.3 & $-67.3$ & 7.3 & 2.7 & 52.5 & 5.6  & 3.1 & 4.3 & 1.5 \\
3 & 5.9 & $-99.8$ & 12.0 & 1.3 & 48.1 & 0.1    & 2.5 & 6.5 & 0.9 \\
4 & 0.4 & $-5.8$  & 6.1 & 0.7 & 69.0 & 0.1    & 0.7 & 0.01 & 3.6 \\

  \hline
\end{tabular}
\begin{flushleft}
\end{flushleft}
\label{table:1614_ave_param}
\end{table*}

CS(1$-$0) line emission traced the dense gas towards HESS\,J1614$-$518. Coverage of the TeV source in the northern and southern halves was provided by MALT$-$45 and Mopra observations respectively. No significant detection in CS(1$-$0) overlapping the TeV source was found in the MALT-45 dataset. However, inspection of the data obtained from Mopra 7\,mm observations revealed a peculiar feature in the narrow velocity range $v_{\text{LSR}} = -47$ to $-44$ km\,s$^{-1}$. No detections in other 7\,mm lines that overlapped HESS\,J1614$-$518 were found in the MALT-45 or Mopra datasets.

\begin{figure*}[!ht]
\centering
\includegraphics[width=0.8\linewidth]{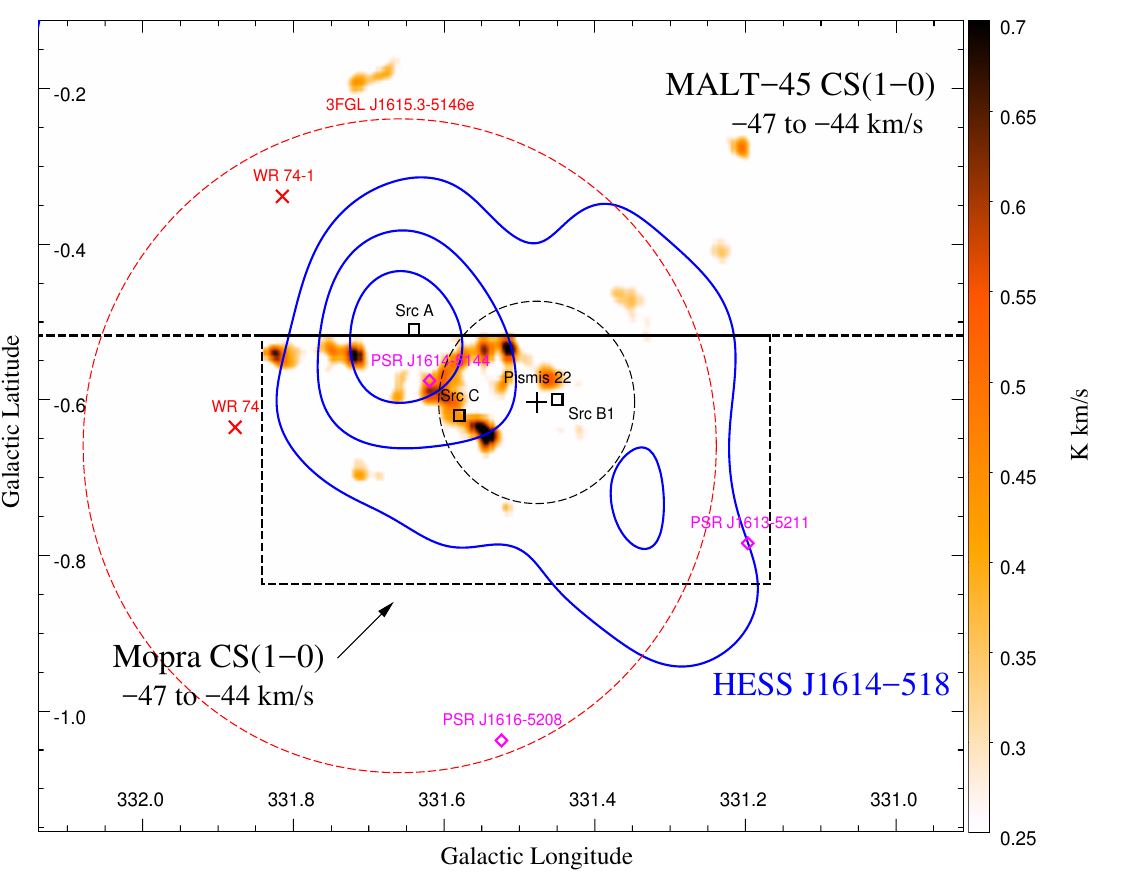}
\caption[what]{CS(1$-$0) integrated intensity image [K\,km\,s$^{-1}$] between $-47$ and $-44$ km\,s$^{-1}$. Overlaid blue contours are HESS excess counts contours towards HESS\,J1614$-$518 at the 30, 45 and 60 levels \citep{hess_plane}. Purple diamonds indicate positions of known pulsars \citep{Manchester2005}. Wolf-Rayet stars WR 74 \citep{2001NewAR..45..135V} and WR\,74-1 \citep{2011AJ....142...40M} are shown as red X's. The 95\% confidence region of the Fermi source 3FGL J1615.3$-$5146e is marked as a dashed red circle \citep{2015ApJS..218...23A}. The centre and extent of the open stellar cluster Pismis 22 is shown as a black plus and dashed circle respectively \citep{2013A&A...558A..53K}. The positions of the X-ray sources Suzaku Src A, XMM-Newton Src B1 and Suzaku Src C are indicated as black squares \citep{2008PASJ...60S.163M,2011PASJ...63S.879S}. The large dashed black rectangle is the extent of the 7mm observations carried out by Mopra. The regions above this rectangle in this image was covered by MALT-45 \citep{Jordan2015}.}
  \label{fig:1614_cs_interest}
\end{figure*}

Figure \ref{fig:1614_cs_interest} is an integrated image of CS(1$-$0) emission in this velocity interval, clearly revealing an open ring of dense gas near the centre of HESS\,J1614$-$518, as well as several dense clumps towards the Galactic-east side of the source. Also displayed in Figure \ref{fig:1614_cs_interest} are the locations of various objects of interest in the region. Several pulsars, Wolf-Rayet stars and X-ray sources are seen towards HESS\,J1614$-$518, with PSR\,J1614-5144 and the X-ray source Suzaku Src C located on the rim of the dense gas ring. The gas ring is seen within the extent of the stellar cluster Pismis 22.

\begin{figure*}[!ht]
\centering
\includegraphics[width=\linewidth]{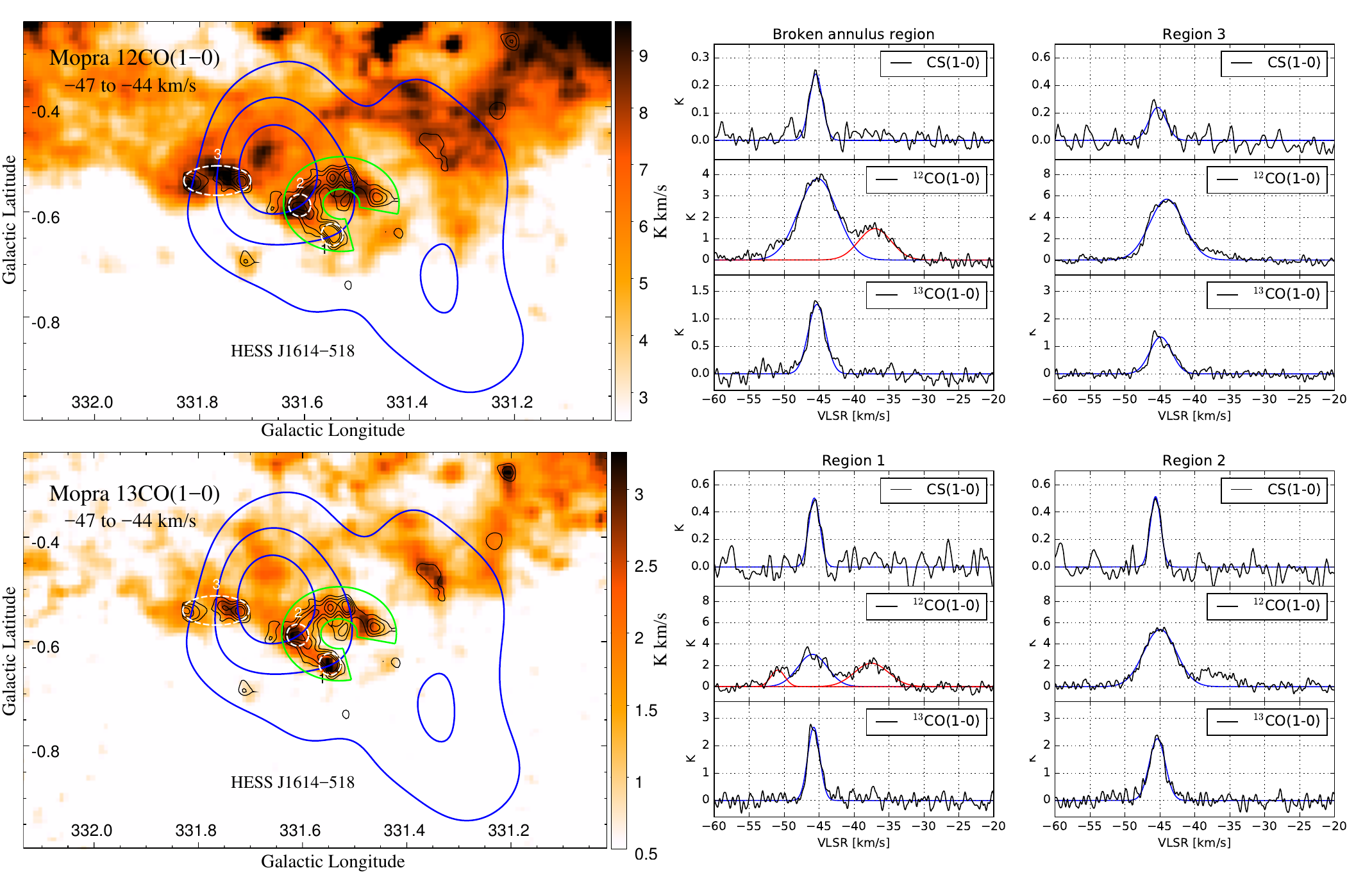}
\caption[what]{\emph{Left:} $^{12}$CO(1$-$0) and $^{13}$CO(1$-$0) integrated intensity images [K\,km\,s$^{-1}$] between $-47$ and $-44$ km\,s$^{-1}$. Overlaid are the CS(1$-$0) emission contours (black) in the same velocity interval and the HESS excess counts contours (blue) \citep{hess_plane}. The solid green broken annulus indicates the region in which an open ring feature is seen clearly in the CS emission. White dashed ellipses indicate additional regions in which spectra were extracted from. \emph{Right:} Solid black lines are the average emission spectra for CS(1$-$0), $^{12}$CO(1$-$0) and $^{13}$CO(1$-$0) within the broken annulus and 3 additional regions indicated in the left panels. The blue lines indicate the Gaussian functions that were used to parametrise the emission. The fit parameters are displayed in Table \ref{table:annulus_param}. The red lines are additional Gaussian fits to unrelated gas components seen at different $v_{\text{LSR}}$.}
  \label{fig:1614_annulus}
\end{figure*}

The panels on the left side of Figure \ref{fig:1614_annulus} are integrated images of $^{12}$CO(1$-$0) and $^{13}$CO(1$-$0) emission in the $v_{\text{LSR}} = -47$ to $-44$ km\,s$^{-1}$ range. The solid green broken annulus encompasses the approximate region in which the open ring in CS appears. The ring feature, while not as pronounced as that seen in CS(1$-$0), can also be seen in this narrow velocity range in the $^{12}$CO(1$-$0) and $^{13}$CO(1$-$0) images.

The panels on the right side of Figure \ref{fig:1614_annulus} show the average spectra of the CS(1$-$0), $^{12}$CO(1$-$0) and $^{13}$CO(1$-$0) within the green broken annulus and the dashed white ellipses. For the broken annulus spectra, an obvious component is seen in all three tracers centred at $v_{\text{LSR}} \sim -45$ km\,s$^{-1}$ (with corresponding kinematic distance $\sim 3.1$ kpc (following \citealt{Brand1993}). The average gas mass and density parameters within the broken annulus were estimated following \S\ref{sec:line_analysis}. These values, and the fitted Gaussian parameters to the spectra, are displayed in Table \ref{table:annulus_param}. For the broken annulus region, the volume chosen for mass and density calculations was a prism formed by the projection of the broken annulus with depth equal to the average annular radius. We note that the calculated CS parameters should be treated as upper limits, due to the choice of the CS/H$_\text{2}$ abundance ratio (see \S\ref{sec:cs_line_analysis}).

\begin{table*}
  \centering
  \caption{Line parameters for the CS(1$-$0), $^{12}$CO(1$-$0) and $^{13}$CO(1$-$0) emission component at $v_{\text{LSR}} \sim -45$ km\,s$^{-1}$ in the broken annulus and additional region apertures indicated in the left panels of Figure \ref{fig:1614_annulus} towards HESS\,J1614$-$518. Gaussian fits to the component seen in the spectra (blue functions in the right panels of Figure \ref{fig:1614_annulus}) were used to find $v_{\text{LSR}}$, $\Delta v_\text{FWHM}$ and $T_{\text{peak}}$. Corresponding calculated gas parameters for the regions are also displayed, which were calculated following \S \ref{sec:line_analysis}. The CS parameters should be treated as upper limits (see \S\ref{sec:cs_line_analysis}). Distance for all calculations has been assumed to be 3.1 kpc. The $^{12}$CO/$^{13}$CO abundance ratio, $X_{12/13}$, at this distance was taken to be 57.0 (see \S\ref{sec:co_line_analysis}).}
  \begin{tabular}{ccccccccc}
  \hline 
  
Region & Tracer & $v_{\text{LSR}}$ & $\Delta v_\text{FWHM}$ & $T_{\text{peak}}$ & Optical & $\overline{N_{H_{2}}}$ & Mass & $\overline{n}$ \\  
& & (km/s) & (km/s) & (K) & depth & ($10^{21}$ cm$^{-2}$) & (M$_{\odot}\times10^{3}$) & ($10^{2}$ cm$^{-3}$) \\  
  \hline 
  
Broken         & CS(1$-$0) & $-45.5$  & 2.4 & 0.2 &       & 9.5 & 8.4 & 11.0 \\
annulus& $^{12}$CO(1$-$0)  & $-45.1$  & 6.4 & 3.6 & 18.1 & 3.9 & 4.9 & 6.4 \\
       & $^{13}$CO(1$-$0)  & $-45.3$  & 2.8 & 1.1 &      & 2.0 & 2.4 & 3.2 \\ 
\\
1      & CS(1$-$0) & $-45.7$  & 1.9 & 0.5  &      & 9.3 & 0.6 & 40.1 \\
& $^{12}$CO(1$-$0) & $-46.0$  & 4.9 & 3.0  & 50.3 & 2.4 & 0.2 & 14.8 \\
& $^{13}$CO(1$-$0) & $-45.8$  & 2.1 & 2.7  &      & 2.9 & 0.3 & 18.6 \\
\\
2      & CS(1$-$0) & $-45.6$  & 1.6 & 0.5  &      & 8.8 & 0.6 & 38.0 \\
& $^{12}$CO(1$-$0) & $-45.0$  & 5.4 & 5.4  & 11.7 & 4.8 & 0.4 & 29.0 \\
& $^{13}$CO(1$-$0) & $-45.4$  & 2.6 & 2.3  &      & 3.2 & 0.3 & 19.5 \\
\\
3      & CS(1$-$0) & $-45.3$  & 2.8 & 0.2  &      & 6.8 & 1.7 & 14.5 \\
& $^{12}$CO(1$-$0) & $-44.0$  & 5.7 & 5.7  &  8.5 & 5.3 & 1.9 & 16.1 \\
& $^{13}$CO(1$-$0) & $-44.9$  & 3.8 & 1.3  &      & 2.7 & 1.0 &  8.3 \\ 

  \hline
\end{tabular}
\label{table:annulus_param}
\end{table*}

The properties of diffuse gas appear to be non-uniform about the open ring. Regions 1 and 2, indicated in Figure \ref{fig:1614_annulus}, are apertures containing two of the brightest regions of the ring.
Their spectra show very similar features in CS(1$-$0) and $^{13}$CO(1$-$0) emission. However, the $^{12}$CO(1$-$0) emission is significantly reduced in region 1 compared with region 2. The gas parameters for these regions are displayed in Table \ref{table:annulus_param}. The contrast between the $^{12}$CO and $^{13}$CO line ratios between regions 1 and 2 are reflected in the $\sim 5 \times$ difference in calculated $^{12}$CO optical depth. The difference in optical thickness about the ring may be caused by variations in the physical properties, such as temperature and density, of the local gas.

Region 3 encloses dense gas clumps seen in CS(1$-$0) in the Galactic-east of HESS\,J1614$-$518. The emission seen in the spectra for this region is similar to that in the open ring regions in that it is centred at the same kinematic velocity of $\sim -45$ km\,s$^{-1}$. The average gas mass and density parameters for this region is also displayed in Table \ref{table:annulus_param}.

\subsubsection*{Column density towards X-ray sources}
X-ray observations towards HESS\,J1614$-$518 with \emph{Suzaku} and \emph{XMM-Newton} have revealed the presence of several X-ray sources as mentioned in \S\ref{sec:intro}. The positions of the \emph{Suzaku} sources Src A, Src B and Src C are shown in Figure \ref{fig:1614_cs_interest}. The X-ray spectrum of Suzaku Src A is well fit by an absorbed power-law model with a hydrogen column density of $N_{\text{H}} = 1.21^{+0.50}_{-0.41} \times 10^{22}$ cm$^{-2}$ \citep{2008PASJ...60S.163M}. The spectrum of XMM-Newton Src B1, thought to be the main component of Suzaku Src B, is described by either an absorbed power-law or an absorbed blackbody model with $N_{\text{H}} = 2.4^{+0.4}_{-0.4} \times 10^{22}$ cm$^{-2}$ and $N_{\text{H}} = 1.1^{+0.3}_{-0.2} \times 10^{22}$ cm$^{-2}$ respectively \citep{2011PASJ...63S.879S}. Suzaku Src C is a late type B star and would not be a counterpart to HESS\,J1614$-$518. 

In order to constrain the distance to Suzaku Src A and XMM-Newton Src B1, we extracted the average spectra from the $^{12}$CO(1$-$0) and H\textsc{i} line data within the X-ray source extents as given in \cite{2008PASJ...60S.163M} and \cite{2011PASJ...63S.879S}. Integrating the emission spectra allows us to calculate $N_{\text{H}}$ by following \S \ref{sec:line_analysis}, and we are then able to find the total cumulative $N_{\text{H}}$ as a function of $v_{\text{lsr}}$. Here, we assume that the gas traced by $^{12}$CO(1$-$0) and H\textsc{i} emission is located at the near distance.

\begin{figure}[!ht]
\centering
\includegraphics[width=\linewidth]{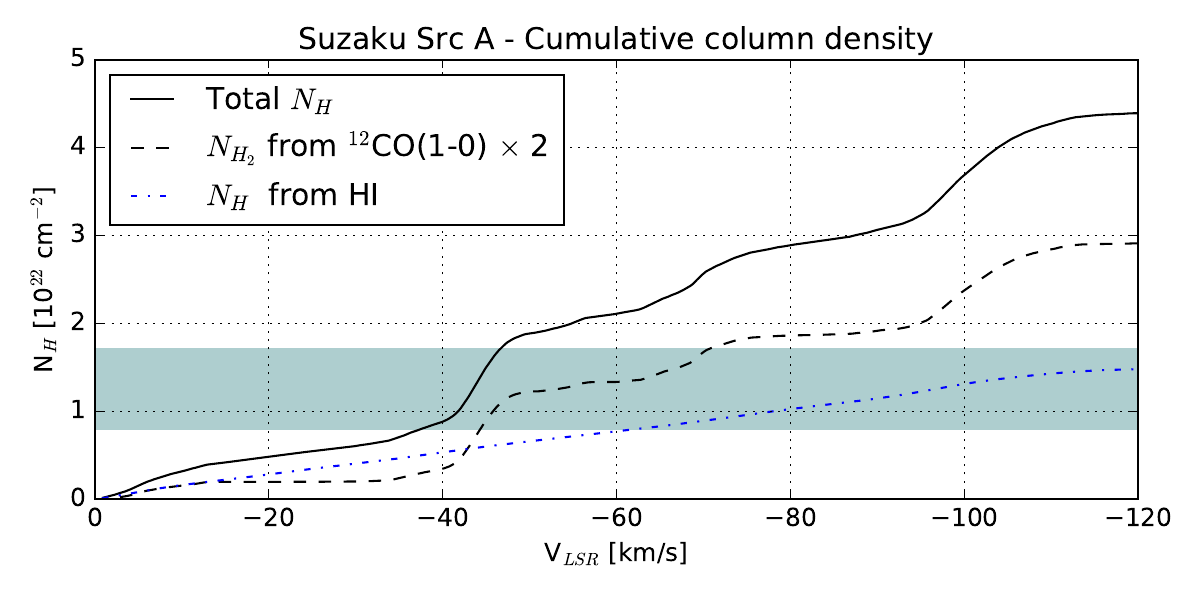}
\includegraphics[width=\linewidth]{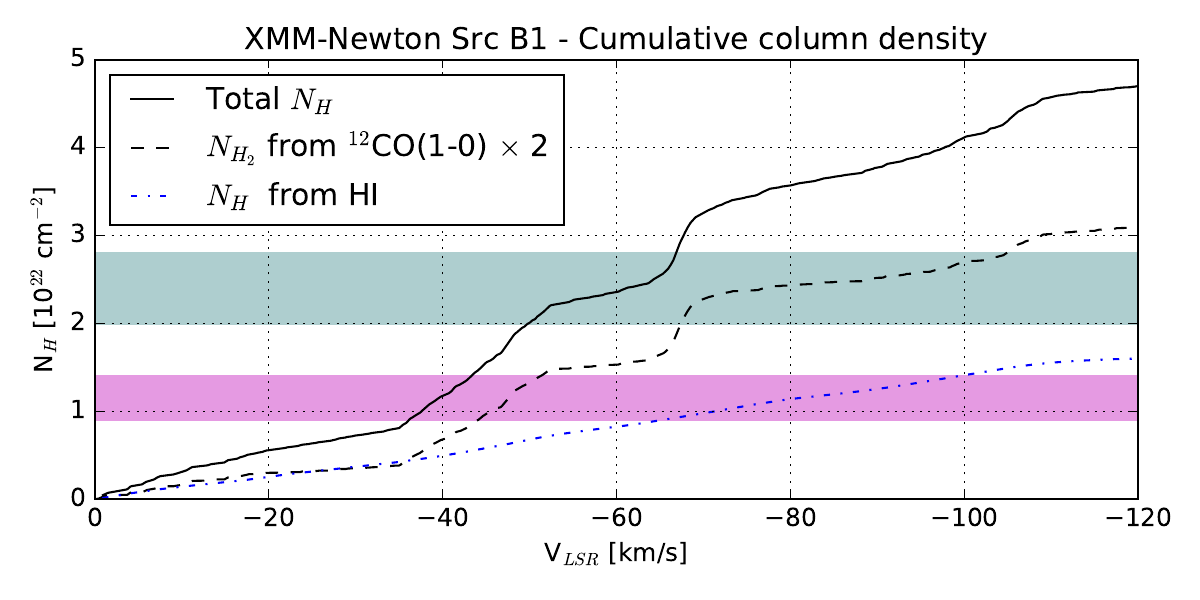}
\caption[what]{Total cumulative hydrogen column density $N_{\text{H}}$ as a function of $v_{\text{LSR}}$ (solid lines) towards the X-rays sources Suzaku Src A (top) and XMM-Newton Src B1 (bottom) which overlap HESS\,J1614$-$518. The cumulative molecular and atomic hydrogen column densities, calculated from CO(1$-$0) and H\textsc{i} data, are shown as dashed and dot-dashed lines respectively. The cyan shaded regions indicate the $N_{\text{H}}$ that were used to fit the spectra of the X-rays sources with absorbed power-law models \citep{2008PASJ...60S.163M,2011PASJ...63S.879S}. The pink shaded region indicates the $N_{\text{H}}$ used in the absorbed blackbody model to fit XMM-Newton Src B1 \citep{2011PASJ...63S.879S}.}
  \label{fig:cumulative_density}
\end{figure}

In Figure \ref{fig:cumulative_density}, we plot the total cumulative $N_{\text{H}}$ towards Suzaku Src A and XMM-Newton Src B1 against $v_{\text{lsr}}$ as a solid black line. We also plot the cumulative $N_{\text{H}_2}$ and $N_{\text{H}}$ that was found using $^{12}$CO(1$-$0) and H\textsc{i} line emission. The shaded regions indicate the values of $N_{\text{H}}$ that was required to the fit the X-ray spectra with an absorbed power-law model (cyan) and an absorbed blackbody model (pink; Src B1 only).

For Src A, the required $N_{\text{H}}$ for the absorbed power-law model occurs in the $v_{\text{lsr}}$ range of $\sim -37$ to $-47$ km\,s$^{-1}$, corresponding to a distance of $\sim 2.6$ to 3.2 kpc \citep{Brand1993}. The $N_{\text{H}}$ needed for the Src B1 absorbed power-law model falls in the $v_{\text{lsr}} \sim -50$ to $-67$ km\,s$^{-1}$ range, with associated kinematic distance of \mbox{$\sim 3.4$} to 4.3 kpc. The $N_{\text{H}}$ requirement for the absorbed blackbody model for Src B1 occurs in the $v_{\text{lsr}} \sim -36$ to $-43$ km\,s$^{-1}$ range, which corresponds to a distance of $\sim 2.6$ to 3.0 kpc.

A distance of 10 kpc was assigned to Suzaku Src\,\,A and XMM-Newton Src B1 \citep{2008PASJ...60S.163M,2011PASJ...63S.879S} based on the comparisons between the best-fit hydrogen column density and the total Galactic H\textsc{i} column density ($\sim 2.2 \times 10^{22}$ cm$^{-2}$, \citealt{Dickey1990}). The blackbody model was assumed for Src\,\,B1 in this case, as it returned similar $N_{\text{H}}$ values as that in the Src\,\,A model. In this section, we have considered the contributions to the total column density from both atomic and molecular hydrogen gas (traced by H\textsc{i} and CO(1$-$0) respectively). This allowed for a more accurate estimation of the column density along the line-of-sight, and hence a better estimate of the distance towards these X-ray sources. If it is the case that Src A is physically related to Src B1, as alluded to in \cite{2011PASJ...63S.879S}, then by assuming the blackbody model for Src B1 we estimate a distance of $\sim 3$ kpc to both sources, based on the hydrogen column density requirements.

\subsection{ISM towards HESS\,J1616$-$508}
\label{sec:result_1616}
\begin{figure}[!ht]
\centering
\includegraphics[width=\linewidth]{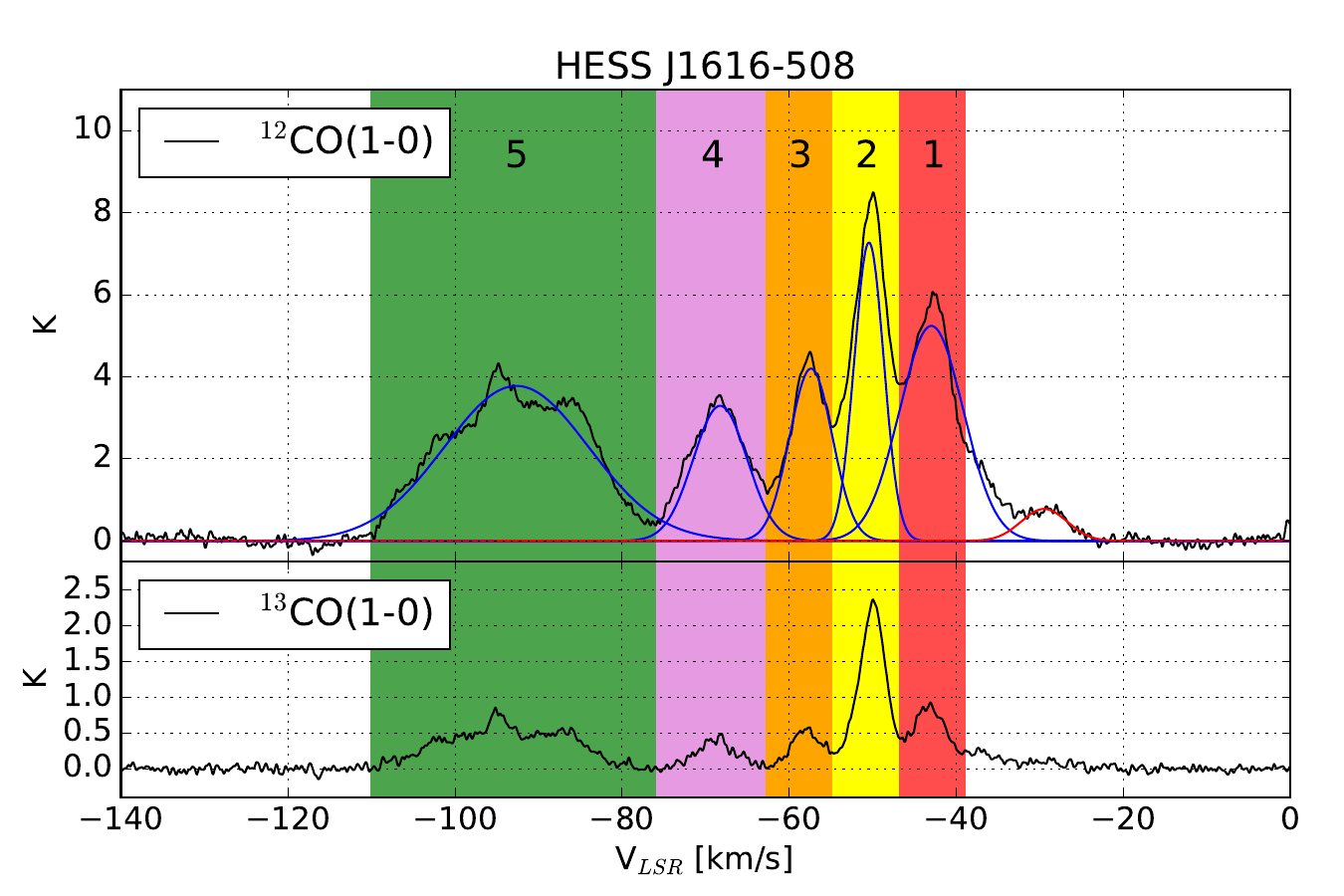}
\caption[what]{The average spectra of $^{12}$CO(1$-$0) and $^{13}$CO(1$-$0) emission within the RMS extent of HESS\,J1616$-$508 as described in \cite{hess_plane}. The velocity intervals of the components used in the integrated images shown in Figure \ref{fig:1616_co} are indicated by the shaded rectangles. Overlaid blue and red lines are the Gaussian fits to the emission, as described in text.}
  \label{fig:1616_co_spectra}
\end{figure}

\begin{figure*}[!ht]
\centering
\vspace{1cm}
\includegraphics[width=\linewidth]{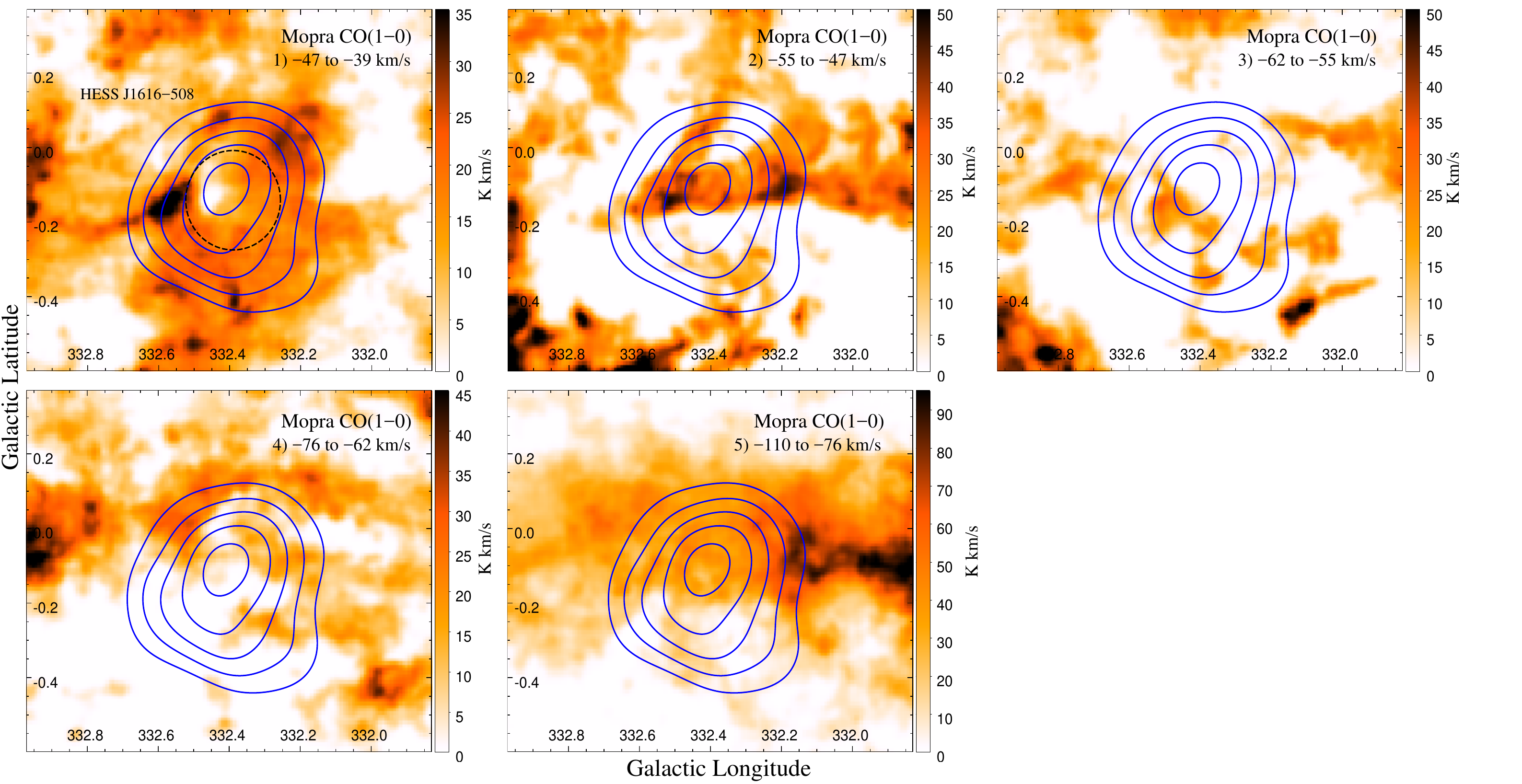}

\vspace{1cm}
\includegraphics[width=\linewidth]{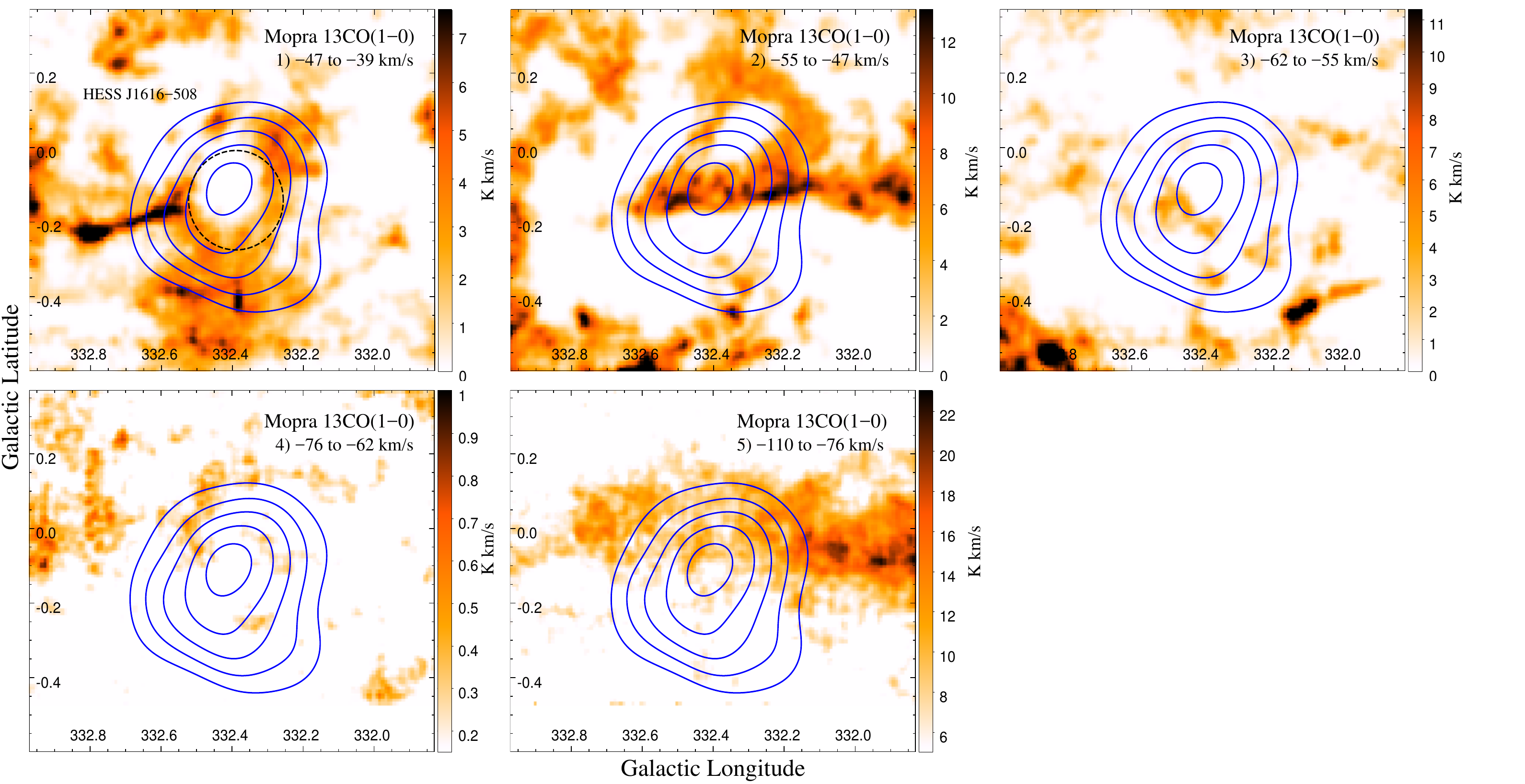}
\caption[what]{Mosaic of $^{12}$CO(1$-$0) (Top) and $^{13}$CO(1$-$0) (Bottom) integrated intensity images [K\,km\,s$^{-1}$] within the labelled velocity intervals towards HESS\,J1616$-$508. Overlaid are HESS excess counts contours (blue) at the 30, 45, 60, 75 and 90 levels, and the dashed black circle in the top left panel of both mosaics is the RMS extent of HESS\,J1616$-$508 \citep{hess_plane}. The average CO(1$-$0) spectra within this region is displayed in Figure \ref{fig:1616_co_spectra}.}
\vspace{1cm}
  \label{fig:1616_co}
\end{figure*}

Figure \ref{fig:1616_co_spectra} displays the average $^{12}$CO(1$-$0) and $^{13}$CO(1$-$0) spectra within the reported RMS extent of HESS\,J1616$-$508 \citep{hess_plane}, which is indicated by the black dashed circles in Figure \ref{fig:1616_co}. Various features along the line of sight in the diffuse gas traced by CO(1$-$0) emission is seen overlapping HESS\,J1616$-$508, with multiple broad components between $-120$ and 0 km\,s$^{-1}$ appearing in the spectra. We have divided the spectra into five velocity components in which emission is prominent. These components are indicated by the shaded rectangles and labelled numerically in Figure \ref{fig:1616_co}.

The features in components 1 through 4 appear to be somewhat blended together, and so we have chosen velocity ranges in which the dominant feature of interest is seen. The spectra for component 5 appears to be composed of several features that overlap and blend together, likely as a result of the line-of-sight being down the tangent of the Norma spiral arm. As we cannot discern an obvious dominant feature, we have chosen a velocity range over all the emission for this component as a first-look approximation. 

A series of Gaussian functions were fit to the $^{12}$CO(1$-$0) spectrum in order to calculate the mass and density parameters of the molecular gas following \S\ref{sec:line_analysis}. The dominant features in components 1 to 4 appeared have have some overlap, yet they are very obviously resolvable. Hence we fit Gaussian functions in order to minimise the effects of cross-contamination. We note that to remove contamination from the extra tail of emission seen towards more positive velocities in component 1, one extra Gaussian function was used in the fitting process. Component 5 appears to consist of at least three or more blended emission features which we cannot resolve. In this case, we roughly approximate the entire emission of component 5 as a single Gaussian function. The fitted Gaussian functions are overlaid on the $^{12}$CO(1$-$0) spectrum in Figure \ref{fig:1616_co_spectra}. Blue functions indicate those that were used in the mass and density calculations for the corresponding labelled components. The red function was the Gaussian used to fit the extra tail of emission near component 1. The parameters of the blue Gaussian functions as well as the calculated gas parameters for each component are displayed in Table\,\,\ref{table:1616_co_param}.

Figure \ref{fig:1616_co} shows a mosaic of the integrated $^{12}$CO(1$-$0) and $^{13}$CO(1$-$0) emission images towards HESS\,J1616$-$508 over the velocity ranges of each component.

In the integrated emission image corresponding to component 1 ($-47$ to $-39$ km\,s$^{-1}$), the molecular gas appears to overlap HESS\,J1616$-$508. The gas extends past the TeV source somewhat in all directions. Interestingly, there appears to be a dip in the $^{12}$CO(1$-$0) emission towards the peak of the TeV emission. The void/dip is much more clearly pronounced in the corresponding integrated $^{13}$CO(1$-$0) emission image. Additionally, there seems to be a thin line of emission extending to the Galactic-east from the peak of the TeV emission which appears more prominently in $^{13}$CO(1$-$0) emission. The intensity of the $^{13}$CO(1$-$0) emission along the thin line appears fairly consistent. However, the intensity of the $^{12}$CO(1$-$0) emission varies, with weaker emission towards the Galactic-east portion of the line, and more intense emission seen in a clump towards the Galactic-west portion. This contrast between the $^{12}$CO and $^{13}$CO line ratios suggests that the $^{12}$CO is more optically thick in the Galactic-east portion of the line feature compared to the clump towards the Galactic-west.

The clump at the Galactic-east end of the line feature is spatially coincident with the H\textsc{ii} region G332.5-00.1 (see Figure \ref{fig:spitzer_hi}). The systematic velocity of the H\textsc{ii} region is $\sim -46$ km\,s$^{-1}$ \citep{1996A&AS..115...81B}, consistent with the velocity of the gas. This suggests that the gas clump in component 1 may be associated with this H\textsc{ii} region.

Component 2 ($-55$ to $-47$ km\,s$^{-1}$) has the strongest feature in the CO spectra. A loop of gas overlaps the TeV source which cuts through the centre of HESS\,J1616$-$508 before looping up and back on itself through the Galactic-north segment.

The morphology of the gas seen in component 3 ($-62$ to $-55$ km\,s$^{-1}$) is patchwork-like, with several regions of gas dispersed mainly in the Galactic-southern parts of HESS\,J1616$-$508.

There is less defined structure in the integrated images of components 4 and 5 ($-76$ to $-62$ km\,s$^{-1}$ and $-110$ to $-76$ km\,s$^{-1}$ respectively). Gas is seen overlapping HESS\,J1616$-$508 in both component 4 and 5, and appear to extend further towards the Galactic-east and Galactic-west. As mentioned before, the spectra for component 5 is very broad ($\sim$ 34 km\,s$^{-1}$) and appears to be composed of several features that overlap and blend together. It is likely that the broad emission here is the result of the sight-line being down the tangent of the Norma spiral arm, and as such there is difficulty in discerning whether these features and physically connected or not.

\begin{table*}
  \centering
  \caption{The parameters of the $^{12}$CO(1$-$0) line emission and the calculated physical parameters for the gas within the indicated RMS extent of HESS\,J1616$-$508 in Figure \ref{fig:1616_co}. Gaussian fits to the spectra were used to determine the line-of-sight velocity, $v_{\text{LSR}}$, line-width (full-width-half-maximum), $\Delta v_\text{FWHM}$, and peak intensity, $T_{\text{peak}}$. The $^{12}$CO/$^{13}$CO abundance ratio, $X_{12/13}$, and optical depth were found following \S\ref{sec:co_line_analysis}. The assumed distances, $d_0$, used in mass and density calculations are the near solutions derived from the Galactic rotation curve presented in \cite{Brand1993}. Calculated mass and density values can be scaled for an arbitrary distance, $d$, using a factor of $(d/d_0)^2$ and $(d/d_0)^{-1}$ respectively.}
  \begin{tabular}{cccccccccc}
  \hline 
  
Component & Distance & $v_{\text{LSR}}$ & $\Delta v_\text{FWHM}$ & $T_{\text{peak}}$ & $X_{12/13}$ & Optical & $\overline{N_{H_{2}}}$ & Mass & $\overline{n}$ \\ 
 & (kpc) & (km/s) & (km/s) & (K) & & depth & ($10^{21}$ cm$^{-2}$) & (M$_{\odot}\times10^{4}$) & ($10^{2}$ cm$^{-3}$) \\
  \hline 
  
1 & 3.0 & $-43.0$ & 8.9 & 5.3 & 57.1 & 8.7  & 7.7 & 2.7 & 7.4 \\

2 & 3.5 & $-50.4$ & 4.2 & 7.4 & 55.3 & 17.0 & 4.9 & 2.3 & 4.2 \\
  
3 & 3.8 & $-57.4$ & 6.1 & 4.3 & 53.0 & 6.6  & 4.1 & 2.3 & 3.1 \\ 

4 & 4.4 & $-68.3$ & 7.5 & 3.4 & 51.9 & 6.4  & 4.1 & 3.0 & 2.7 \\ 

5 & 5.6 & $-92.6$ & 20.3 & 3.8 & 48.4 & 7.8  & 12.4 & 14.8 & 6.5 \\ 

  \hline
\end{tabular}
\label{table:1616_co_param}
\end{table*}

\begin{figure*}[!ht]
\centering
\includegraphics[width=0.9\linewidth]{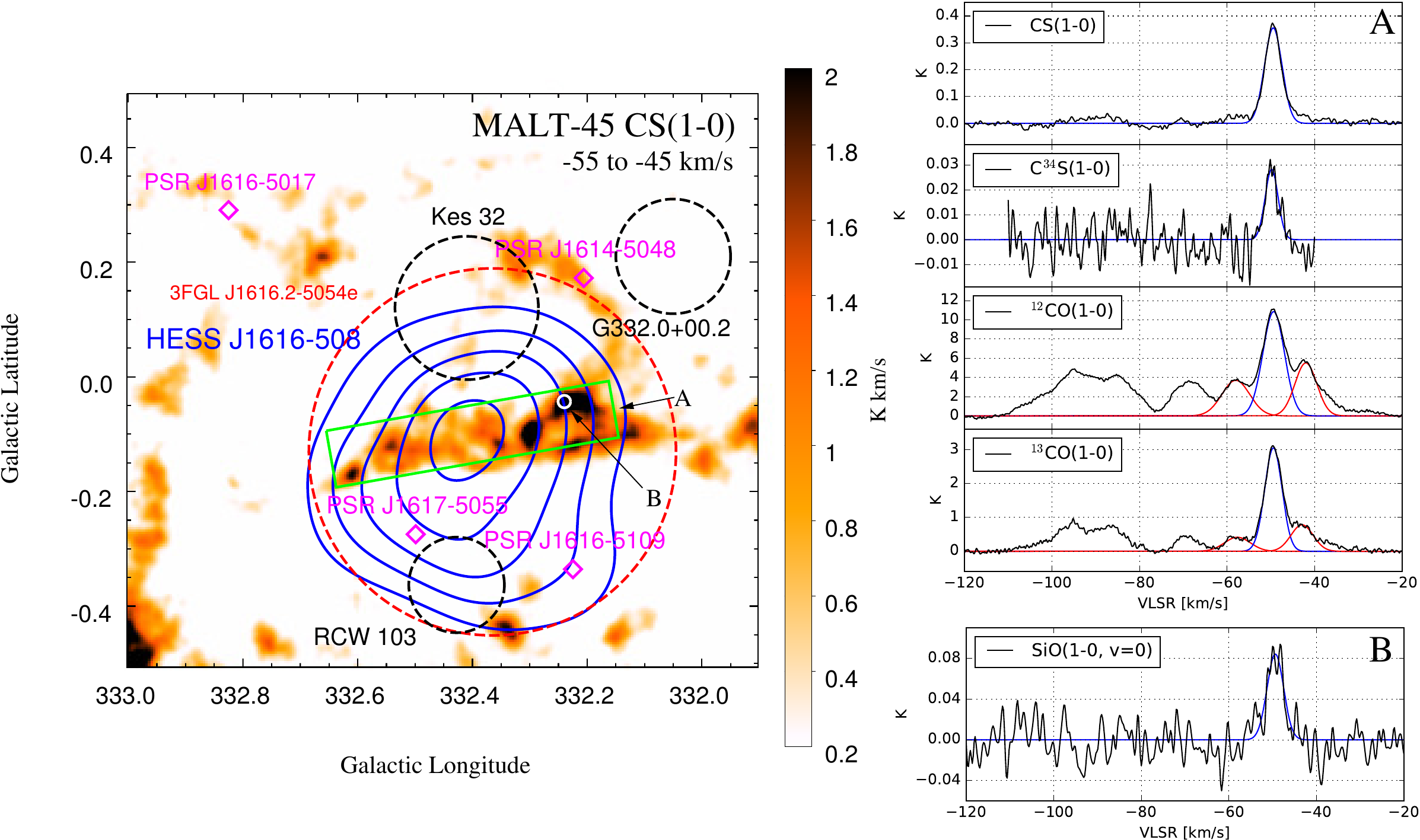}
\caption[what]{\emph{Left:} CS(1$-$0) integrated intensity image [K\,km\,s$^{-1}$] between $-55$ and $-45$ km\,s$^{-1}$ towards HESS\,J1616$-$508. Overlaid blue contours are HESS excess counts contours at the 30, 45, 60, 75 and 90 levels \citep{hess_plane}. The 95\% confidence region of the Fermi source 3FGL J1616.2$-$5054e is marked as a dashed red circle \citep{2015ApJS..218...23A}. Purple diamonds indicate positions of known pulsars \citep{Manchester2005}, and black dashed circles indicate extent of known SNRs in the region \citep{2014BASI...42...47G}. The solid green rectangle indicates the region of the loop feature seen in CS(1$-$0) emission that cuts horizontally through the TeV source. \mbox{\emph{Top right:}} Average emission spectra (black) for CS(1$-$0), C$^{34}$S(1$-$0), $^{12}$CO(1$-$0) and $^{13}$CO(1$-$0) within the green rectangular region (region `A') indicated in the left panel. Blue lines indicate Gaussian functions used to parametrise the emission seen in the  $\sim -55$ to $-45$ km\,s$^{-1}$ velocity range. Fit parameters are displayed in Table \ref{table:1616_loop_param}. Red lines indicate Gaussian fits to other gas components seen in nearby velocities. \mbox{\emph{Bottom-right:}} Average SiO(1$-$0, v=0) spectrum in the small circular aperture labelled `B' in the left panel.}
  \label{fig:1616_cs}
\end{figure*}

Dense gas was traced towards HESS\,J1616$-$508 using CS(1$-$0) emission from the MALT-45 7\,mm survey. CS(1$-$0) emission is seen overlapping the TeV source in the velocity range $\sim -55$ to $-45$ km\,s$^{-1}$, which corresponds to component 2 in the CO(1$-$0) emission. The left panel of Figure \ref{fig:1616_cs} is a integrated image of CS(1$-$0) emission over said velocity ranges. The morphology of the emission matches very well with the CO(1$-$0) emission in component 2. This suggests that the CS is tracing the dense molecular gas embedded in the inner region of the loop structure traced in CO.

A large fraction of the loop feature cuts across HESS\,J1616$-$508. The solid green rectangle in the left panel of Figure \ref{fig:1616_cs} (labelled `A') is the approximate region of where CS(1$-$0) emission in this loop overlaps the TeV source. Detection in the isotopologue transition C$^{34}$S(1$-$0) was also made within this region. The top right panel of Figure \ref{fig:1616_cs} shows the average spectra of the CS(1$-$0), C$^{34}$S(1$-$0), $^{12}$CO(1$-$0) and $^{13}$CO(1$-$0) emission extracted from the green rectangular aperture. A significant feature is seen in all three tracers centred at $v_{\text{LSR}} \sim -49$ km\,s$^{-1}$, which corresponds to a kinematic distance of $\sim 3.4$ kpc (following \citealt{Brand1993}). Using these spectra, the average gas mass and density parameters for this region were calculated following \S\ref{sec:line_analysis}. The Gaussian function parameters used to fit the spectra and the calculated gas parameters are displayed in Table \ref{table:1616_loop_param}. For these mass and density calculations, the geometry of the assumed volume was a prism with depth equal to the smaller side of the rectangle. The calculated CS parameters should be treated as upper limits, due to the choice of the CS/H$_\text{2}$ abundance ratio (see \S\ref{sec:cs_line_analysis}).

Emission in the SiO(1$-$0, v=0) line was found in a small region labelled `B' in the left panel of Figure \ref{fig:1616_cs} at the same velocity as the loop feature ($\sim$ -49 km\,s$^{-1}$). The average SiO spectrum within this region is shown in the bottom-right panel of Figure \ref{fig:1616_cs}.

An IR dark cloud is seen in the \textit{Spitzer} GLIMPSE data with similar morphology to the `hook' region of the gas loop feature towards the Galactic North-West of HESS\,J1616$-$508. IR emission in the `bar' region of the gas loop that cuts through the TeV source suggest regions of star formation activity. The SiO emission in region B indicates a shocked region, likely due to recent star formation. A more detailed analysis and discussion of this gas loop feature focusing particularly on the star formation activity will be presented in an upcoming paper (Romano et al. 2017 \textit{in prep}).

\begin{table*}
  \centering
  \caption{CS(1$-$0), C$^{34}$S(1$-$0), $^{12}$CO(1$-$0) and $^{13}$CO(1$-$0) line parameters for the emission feature at $v_{\text{LSR}} \sim -49$ km\,s$^{-1}$ within the green rectangular aperture indicated in the left panel of Figure \ref{fig:1616_cs}. Gaussian fits to the component seen in the spectra (blue functions in the right panels of Figure \ref{fig:1616_cs}) were used to find $v_{\text{LSR}}$, $\Delta v_\text{FWHM}$ and $T_{\text{peak}}$. Corresponding calculated gas parameters for the region are also displayed, calculated following \S \ref{sec:line_analysis}. The CS parameters should be treated as upper limits (see \S\ref{sec:cs_line_analysis}). Distance for the calculations has been assumed to be 3.4 kpc. Following \S\ref{sec:co_line_analysis}, the $^{12}$CO/$^{13}$CO abundance ratio, $X_{12/13}$, at this distance was taken to be 55.5.}
  \begin{tabular}{cccccccc}
  \hline 
  
Tracer & $v_{\text{LSR}}$ & $\Delta v_\text{FWHM}$ & $T_{\text{peak}}$ & Optical & $\overline{N_{H_{2}}}$ & Mass & $\overline{n}$ \\  
& (km/s) & (km/s) & (K) & depth & ($10^{21}$ cm$^{-2}$) & (M$_{\odot}\times10^{4}$) & ($10^{2}$ cm$^{-3}$) \\  
  \hline 
  
       CS(1$-$0) & $-49.5$  & 4.7 & 0.36  & 1.3  & 29.4 & 8.4  & 18.9 \\
C$^{34}$S(1$-$0) & $-49.8$  & 3.5 & 0.028 &      &      &      &      \\
$^{12}$CO(1$-$0) & $-49.3$  & 5.9 & 10.9  & 12.5 & 10.1 & 4.0  & 9.1 \\
$^{13}$CO(1$-$0) & $-49.4$  & 4.7 & 3.0   &      & 7.4  & 3.0  & 6.7 \\ 
  \hline
\end{tabular}
\label{table:1616_loop_param}
\end{table*}

\subsection*{HI data}
\label{sec:results_HI}
H\textsc{i} data was available towards HESS\,J1614$-$518 and HESS\,J1616$-$508 from the SGPS \citep{SGPS} which was used to study the distribution of atomic gas towards the TeV sources. Using the same velocity intervals where components of emission were seen in the CO data (see Figures \ref{fig:1614_co} and \ref{fig:1616_co}), integrated H\textsc{i} maps were generated. These integrated maps can be seen in the appendix Figures\,\ref{fig:SGPS_1614} and \ref{fig:SGPS_1616}. Also displayed in these figures are the average H\textsc{i} emission spectra within the extents of HESS\,J1614$-$518 and HESS\,J1616$-$508.

The integrated H\textsc{i} images towards HESS\,J1614$-$518 show no obvious morphological features overlapping or anti-correlating the TeV source. In the integrated H\textsc{i} image towards HESS\,J1616$-$508 for component 1 ($-47$ to $-39$ km\,s$^{-1}$), a relative increase in the amount of emission is seen towards the central peak of the TeV source. This is anti-correlated with the void observed in CO(1$-$0) emission in the same velocity interval. There also appears to be a localised region of diminished emission to the Galactic-east of the TeV peak, which is coincident with the clump at the end of the line feature seen in CO (Figure \ref{fig:1616_co}, see \S\ref{sec:result_1616}). In component 5 ($-110$ to $-76$ km\,s$^{-1}$), a region of more intense H\textsc{i} emission is seen in the Galactic-southeast portion of HESS\,J1614$-$518. This feature does not appear in the molecular gas data.

A broken ring-like feature of diminished H\textsc{i} brightness temperature is seen towards the northern part of HESS\,J1616$-$508 in all of integrated images except in component 4 ($-76$ to $-62$ km\,s$^{-1}$). This feature is likely associated with SNR Kes 32 as it is both positionally coincident and morphologically similar. Inspection of the H\textsc{i} data shows that this feature appears in several negative velocities up to $\sim -90$ km\,s$^{-1}$. If this H\textsc{i} feature is associated with Kes 32, it would imply a distance of at least $\sim 5.3$ kpc to the SNR (using the rotation curve in \citealt{Brand1993}).

\subsection*{HEAT [CI] data}
Data in the atomic carbon [CI] (J=2-1) line towards HESS\,J1616$-$508 was available in the second data release (DR2) from the High Elevation Antarctic Terahertz (HEAT) telescope\footnote{http://soral.as.arizona.edu/heat/}.

While the standard tracer of molecular hydrogen gas is CO, the abundance of the CO can be greatly reduced due to photo-dissociation by far-UV radiation and interactions with CRs (e.g. \citealt{2013ARA&A..51..207B}) in the outer envelopes of molecular cloud structures. The carbon in these outer envelopes will then exist as C or C$^+$. Emission from neutral atomic carbon, [CI], generally occurs in gas where molecular hydrogen exists without significant CO, and thus [CI] is a good tracer for molecular gas that is ``dark" to standard survey techniques (e.g. \citealt{2015ApJ...811...13B} and references within).

Figure \ref{fig:HEAT} in the appendix displays a series of integrated [CI] emission images in the same velocity intervals as that in Figure \ref{fig:1616_co}. The bottom-right panel of Figure \ref{fig:HEAT} shows the [CI] spectra (in blue) within the reported RMS extend of HESS\,J1616$-$508.

The beamsize of HEAT [CI] (2$'$.5) is relatively larger compared with that of Mopra CO (0$'$.6). However, while intricate comparisons between the [CI] and CO are unavailable, the same general distribution and morphology is seen in both HEAT [CI] and Mopra CO data. In particular, the void feature towards the peak of the TeV source and the region of more intense emission extending to the Galactic-east, seen in component 1 of the CO emission can also be identified in the [CI]. Similarly, the extended loop feature in component 2 of the CO emission is also clearly visible in the [CI]. There are no obvious discrepancies between the [CI] and CO emission.

Future investigation using the data may be able to reveal the fraction of ``dark'' gas traced by [CI] that is not seen by the conventional CO tracers.

\section{Discussion}
The mechanisms behind the production of TeV $\gamma$-rays seen from HESS\,J1614$-$518 and HESS\,J1616$-$508 are unclear, as alluded to in \S\ref{sec:intro}, and are a key question in unravelling the nature of these sources. In the following section, we discuss the implications of our interstellar gas study on the different origin scenarios of HESS\,J1614$-$518 and HESS\,J1616$-$508. The same general methods employed here were applied to another pair of TeV $\gamma$-ray sources, HESS\,J1640$-$465 and HESS\,J1641$-$463, in \cite{2017MNRAS.464.3757L}.

\subsection{Hadronic production of TeV $\gamma$-rays}
Gas that was traced in this study along the line of sight towards HESS\,J1614$-$518 and HESS\,J1616$-$508 may be acting as potential targets for accelerated CRs, producing TeV $\gamma$-rays in hadronic p-p interactions. In \cite{1991_Aharonian}, a relationship between the flux of $\gamma$-rays and the mass and distance of the target material was derived. Above some energy $E_{\gamma}$, the expected $\gamma$-ray flux is given by:

\begin{equation}
\label{eqn:gamma_from_gmc}
F(\geq E_{\gamma}) = 2.85\times10^{-13} E^{-1.6}_{\gamma} \left(\dfrac{M_{5}}{d^{2}_{kpc}}\right)k_{CR} \quad \mathrm{cm^{-2}s^{-1}}
\end{equation}
where the CR-target material mass, $M_{5}$, is given in units of $10^{5}$ M$_{\odot}$, the distance to the material, $d_{kpc}$, is given in kpc, the minimum energy of $\gamma$-rays, $E_{\gamma}$, is given in TeV, and where the parameter $k_{CR}$ is the CR enhancement factor above that observed at Earth. The above equation assumes that the target ISM is located some distance from the CR source. The spectrum of CRs, an $E^{-1}$ integral power law at the accelerator, would have steepened to an $E^{-1.6}$ spectrum due to diffusion in transport.

From \mbox{\cite{hess_plane}}, the $\gamma$-ray flux is $F(\geq\text{200 GeV}) = 57.8 \times 10^{-12}$ cm$^{-2}$\,s$^{-1}$ for HESS\,J1614$-$518 , and $F(\geq\text{200 GeV}) = 43.3 \times 10^{-12}$ cm$^{-2}$\,s$^{-1}$ for HESS\,J1616$-$508.

The sum of the molecular and atomic gas, as traced by CO(1$-$0) and H\textsc{i} respectively, was taken to be the total amount of CR-ray target material. We use this total mass of atomic and molecular gas seen in each `component', as described in \S\ref{sec:results}, to calculate the required CR enhancement factors $k_{CR}$. We note that $k_{CR}$ is effectively independent of any distance assumptions as the distance term of Equation \ref{eqn:gamma_from_gmc} is cancelled out by the distance terms used in the mass calculations. The calculated $k_{CR}$ values are displayed in Table \ref{table:Kcr}, and are discussed in more detail in \S\ref{sec:discussion_1614} and \S\ref{sec:discussion_1616}. We find that the dominant contribution ($\sim 70 - 90\%$) to the total gas mass in each component is from the molecular gas portion. We note that the $k_{CR}$ values presented are applicable to $\gamma$-rays with energy $E_\gamma >$ 200 GeV, corresponding to CRs with energies $E_p \gtrsim$ 1 TeV. As such, any CR energetics are to be treated as lower limits on the total CR energy.

\begin{table*}
  \centering
   \caption{Cosmic-ray enhancement values, k$_{\text{CR}}$, within the extents of HESS\,J16414-518 and HESS\,J1616-508, for the gas components discussed in \S\ref{sec:results} (see Figures \ref{fig:1614_co} and \ref{fig:1616_co}). The molecular gas masses come from CO analysis, while the atomic gas masses are from H\textsc{i} analysis. The total number density $\overline{n}$ includes both molecular and atomic gas. Note that k$_{\text{CR}}$ is independent of the assumed distance, as described in text.}
\begin{tabular}{ccccccccc}
\hline 
Region & $v_{\text{LSR}}$ & Distance & Molecular mass & Atomic mass & Total mass & $\overline{n}$ & k$_{\text{CR}}$ \\ 
• & (km/s) & (kpc) & (M$_{\odot}$) & (M$_{\odot}$) & (M$_{\odot}$) & (10$^2$ cm$^{-3}$) & • \\ 
\hline 
HESS\,J1614$-$518 & $-50$ to $-40$  & 3.1 & 19,000 & 4,100  & 23,000 & 2.2 & 650 \\ 
                • & $-75$ to $-60$  & 4.3 & 43,000 & 11,000 & 54,000 & 1.9 & 530 \\ 
               • & $-115$ to $-90$  & 5.9 & 65,000 & 21,000 & 86,000 & 1.2 & 630 \\ 
                • & $-15$ to $0$    & 0.4 &    100 &    100 &    200 & 7.8 & 1240 \\ 
\\
HESS\,J1616$-$508 & $-47$ to $-39$  & 3.0 & 27,000 & 2,000 & 29,000 & 7.9 & 360 \\ 
              • & $-55$ to $-47$  & 3.5 & 23,000 & 2,500 & 26,000 & 4.7 & 550 \\ 
              • & $-62$ to $-55$  & 3.8 & 23,000 & 2,400 & 25,000 & 3.4 & 670 \\ 
              • & $-76$ to $-62$  & 4.4 & 30,000 & 6,300 & 36,000 & 3.2 & 620 \\ 
              • & $-110$ to $-76$ & 5.6 & 150,000 & 23,000 & 170,000 & 7.4 & 210 \\ 
\hline 
\end{tabular} 
  \centering
\label{table:Kcr}
\begin{flushleft}

\end{flushleft}
\end{table*}

If we consider a hadronic scenario for HESS\,J1614$-$518 and HESS\,J1616$-$508, the total CR energy budget, $W_p$, can be expressed by the relation \mbox{$W_p = L_{\gamma} \tau_{pp}$}, where $L_{\gamma}$ is the $\gamma$-ray luminosity. The cooling time of CR protons, $\tau_{pp}$, via proton-proton collisions, can be given by the expression $\tau_{pp} \approx 6 \times 10^7 (n/1\,\text{cm}^{-3})^{-1}$ yr \citep{1996A&A...309..917A}, where $n$ is the number density of the target gas.

We use the number densities of the gas $\overline{n}$ in the gas components seen towards HESS\,J1614$-$518 and HESS\,J1616$-$508 (see Tables \ref{table:1614_ave_param}, \ref{table:1616_co_param} and \ref{table:Kcr}). The value of $\overline{n}$ is of order $\sim10^{2}$ cm$^{-3}$ for HESS\,J1614$-$518 and of order $\sim10^{2}$ to $10^{3}$ cm$^{-3}$ for HESS\,J1616$-$508. We calculate the $\gamma$-ray luminosity above 200 GeV of HESS\,J1614$-$518 and HESS\,J1616$-$508 at each of the assumed distances to the gas components. $W_p$ is then $\sim10^{48}$ erg for HESS\,J1614$-$518 and $\sim10^{47}$ to $10^{48}$ erg for HESS\,J1616$-$508, which are reasonable fractions of the canonical amount of energy which is injected by a SNR into CRs ($\sim10^{50}$ erg).

\subsection{Leptonic production of TeV $\gamma$-rays}
The leptonic production of TeV $\gamma$-rays involves multi-TeV electrons and their interactions via the inverse-Compton effect with ambient background photons. In the case where a potential accelerator is spatially offset from the TeV emission, CR electrons may be travelling across the intervening ISM diffusively. Within the molecular clouds, the magnetic field strength is typically enhanced \citep{2010ApJ...725..466C} and as a result CR electrons suffer heavy synchrotron radiation losses.

The synchrotron cooling time of CR electrons can be given by $\tau_{\text{sync}} \approx (b_{s}\gamma_e)^{-1}$ s, where $b_{s} = 1.292 \times 10^{-15} (B/\text{mG})^{2}$ s$^{-1}$ is dependent on the local magnetic field strength $B$, and $\gamma_e $ is the Lorentz factor of the electron. We calculate the magnetic field strength of the ISM using the values of $\overline{n}$ from our gas analyses and following \cite{2010ApJ...725..466C}. Over a distance $d$ from the injector, the diffusion time of CR electrons  is given by $\tau_{\text{diff}} = d^{2}/(6D(E))$, where $D(E)$ is the diffusion coefficient for charged particles with energy $E$ \citep{1964ocr..book.....G}. $D(E)$ for CR protons and electrons can be found by using equation (2) from \cite{2007Ap&SS.309..365G}:
\begin{equation}
D(E) = \chi D_0 \left(\dfrac{E/\text{GeV}}{B/3\,\mu\text{G}}\right)^{\delta}
\end{equation}
where $D_0 = 3 \times 10^{27}$ cm$^{2}$\,s$^{-1}$ and $\delta = 0.5$. We assumed a suppression factor $\chi = 0.1$, consistent with values adopted in previous studies of the ISM towards the TeV $\gamma$-ray source W28 \citep{2010A&A...516L..11G,2010sf2a.conf..313G,2012MNRAS.421..935L}.

Where appropriate in the discussion below of possible source associations with HESS\,J1614$-$518 and HESS\,J1616$-$508, we consider the synchrotron cooling and diffusion time-scales to assess the plausibility of potential leptonic scenarios. The HESS analyses for HESS\,J1614$-$518 and HESS\,J1616$-$508 were performed for all observed events. Due to the power-law nature of the $\gamma$-ray spectrum from these sources, most of the detected photons have energies around the HESS detection threshold of $\sim 200$ GeV. As such, when discussing leptonic scenarios, we consider electron energies of \mbox{$E_{e}$ = 5 TeV}, as inverse-Compton scattering would produce $\gamma$-rays with energies near the lower limit of detectability by HESS.

In the following sections, we will discuss the hadronic and leptonic scenarios for both HESS\,J1614$-$518 and HESS\,J1616$-$508.

\subsection{HESS\,J1614$-$518}
\label{sec:discussion_1614}

We discuss now the implications that the gas data considered in this study have on possible TeV $\gamma$-ray production scenarios for HESS\,J1614$-$518 that have been previously suggested in the literature.

Recent preliminary results from HESS suggest that HESS\,J1614$-$518 has a shell-like TeV $\gamma$-ray morphology and may be a SNR candidate \citep{doi:10.1063/1.4968934}. While there are currently no known SNRs towards HESS\,J1614$-$518, it may be possible that an undetected SNR is present, responsible for the TeV $\gamma$-rays by accelerating CRs that interact with the ISM. It has previously been postulated that the X-ray source XMM-Newton Src B1 is an Anomalous X-ray Pulsar (APX), with Suzaku Src A being a shocked region of the related SNR \citep{2011PASJ...63S.879S}.

This SNR scenario was further investigated by \cite{2011ApJ...740...78M}. By assuming a power-law distribution of a population of accelerated protons and a source distance of 10 kpc, an ambient matter density of 100 cm$^{-3}$ was required to reproduce the $\gamma$-ray spectrum of HESS\,J1614$-$518 via a hadronic interaction model. The distance assumption was taken from those assigned to Suzaku Src A and XMM-Newton Src B1 \citep{2008PASJ...60S.163M,2011PASJ...63S.879S}, based on the best-fit hydrogen column density. However, this estimate was made only by comparison with the total Galactic H\textsc{i} column density. By considering the contribution to the column density from both atomic and molecular gas in \S\ref{sec:result_1614}, we estimate a distance of $\sim$ 3 kpc to both X-ray sources. If we apply this distance to the SNR scenario, scaling the results of \cite{2011ApJ...740...78M} would imply an ambient matter density of $\sim10$ cm$^{-3}$ would be required.

In \S\ref{sec:result_1614}, CO(1$-$0) observations revealed four components where the diffuse gas is seen towards HESS\,J1614$-$518. According to Table \ref{table:Kcr}, the total number density of the gas in each of these components is of the order 10$^2$ cm$^{-3}$, which would satisfy the requirements from \cite{2011ApJ...740...78M}. In terms of morphology, there is no strong correspondence that is immediately obvious in any of the components. However, Component\,\,1 would be the most likely candidate for association as the diffuse gas traced by $^{12}$CO(1$-$0) overlaps most of of HESS\,J1614$-$518, with gas traced by $^{13}$CO(1$-$0) appearing towards the TeV peak. This component is also located at a distance of \mbox{$\sim 3$ kpc}, which is consistent with our estimated distance to Suzaku Src\,\,A, XMM-Newton Src B1, and consequently the SNR in this scenario.

The required $k_{\text{CR}}$ value for the total gas mass in component 1 is $\sim 650$ (see Table \ref{table:Kcr}) and is consistent with a young SNR ($\lesssim5$ kyr) injecting accelerated CRs into the local ISM \citep{1996A&A...309..917A}. At a distance of 3 kpc, the distance between Suzaku Src A and XMM-Newton Src B1 is calculated to be $\sim 11$ pc. If the radius of the SNR is the same, the age of the SNR can be roughly estimated to be $\sim 3000$ yr using typical typical values for SN explosions (injection energy $E \sim 10^{51}$\,erg, density $n_\text{H} \sim 1$ cm$^{-3}$) and the Sedov-Taylor solution \citep{1959sdmm.book.....S,1950RSPSA.201..159T}. Thus it is possible that the gas in component 1 is the target material for the accelerated particles from this SNR.

In order to have such an SNR, a high mass progenitor star is required. These stars typically have large stellar winds which can blow out cavities within the ISM. The presence of the dense broken-ring of gas discussed in \S\ref{sec:result_1614}, located at the same distance of $\sim$ 3 kpc, may be indicative of this scenario, where dense compressed gas has been swept up by the stellar winds of the progenitor star. To create this wind blown ring of gas, an O or B type progenitor star with mass $\sim 20$ M$_{\bigodot}$ would be required, based on an estimated radius of $\sim$ 11 pc \citep{2013ApJ...769L..16C}. Following the model presented in \cite{1999ApJ...511..798C}, and using the parameters of the progenitor stars from \cite{2013ApJ...769L..16C}, the total energy input by the stellar winds would be $\sim 5 \times 10^{48}$ erg in order to produce the $\sim11$ pc radius ring.

A SNR may be a somewhat plausible scenario in this sense, supported by recent preliminary results suggesting a shell-like TeV $\gamma$-ray morphology of HESS\,J1614$-$518 \citep{doi:10.1063/1.4968934}. However, no evidence of an SNR has yet been detected at this position towards HESS\,J1614$-$518, and it may be that this is a SNR only seen in TeV $\gamma$-rays.

The stellar winds from stars in the open stellar cluster Pismis 22 have also been considered as a possible association with HESS\,J1614$-$518 \citep{2008AIPC.1085..241R}. \cite{2011ApJ...740...78M} calculated that, based on energetics requirements and an estimated cluster age of 40\,Myr \citep{Piatti}, the stellar winds from two O-type stars were required to produce the TeV $\gamma$-rays in this scenario. However, this scenario assumed a molecular cloud of ambient density 100 cm$^{-3}$ which, the authors noted, had not been found in previous investigations of the ISM using Nanten data \citep{2008AIPC.1085..241R}. 

The estimated distance to the cluster is $\sim 1$ to 2 kpc \citep{Piatti,2013A&A...558A..53K}, with corresponding $v_{\text{LSR}} \sim -13$ to $-27$ km\,s$^{-1}$ using the rotation curve from \cite{Brand1993}. We find no evidence of a molecular cloud at the current estimated distances to Pismis 22, as there were no components of gas detected at the associated velocities in our ISM study. However, the stellar wind scenario may still be possible if the distance to Pismis 22 is underestimated, and if the responsible stars are instead located within the components of gas observed overlapping HESS\,J1614$-$518. 

\begin{figure}[!ht]
\centering
\includegraphics[width=1.0\linewidth]{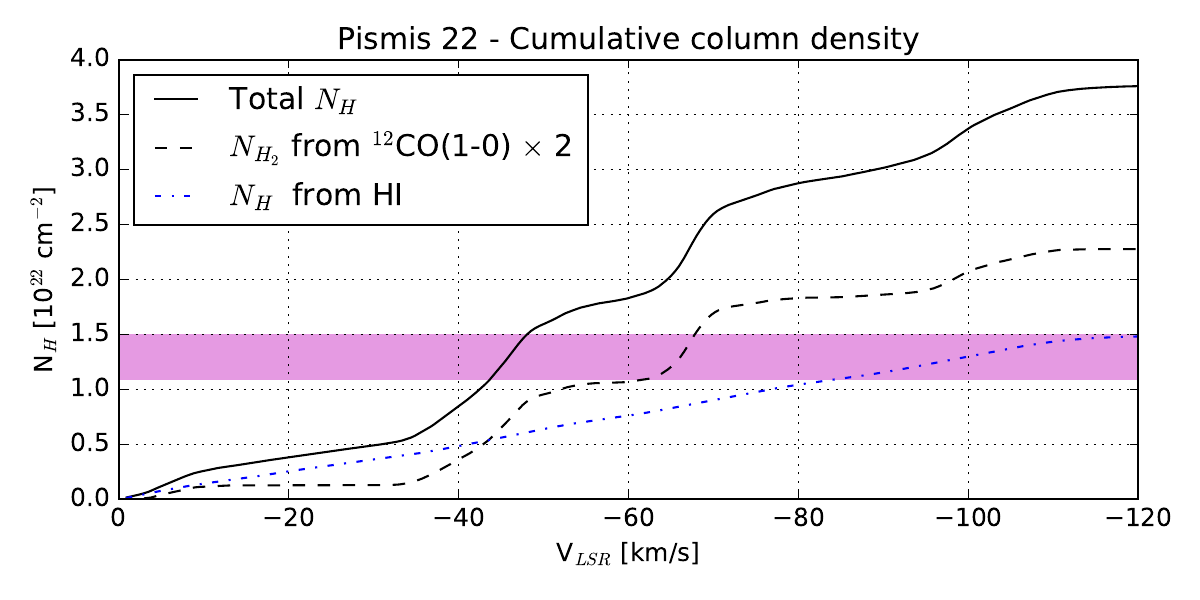}
\caption[what]{Total cumulative hydrogen column density $N_{\text{H}}$ as a function of $v_{\text{LSR}}$ (solid lines) towards the stellar cluster Pismis 22. The cumulative molecular and atomic hydrogen column densities, calculated from $^{12}$CO(1$-$0) and H\textsc{i} data, are shown as dashed and dot-dashed lines respectively. The shaded region indicates the $N_\text{H}$ required to to obtain the value of $A_v$ used in \cite{Piatti}, which was calculated using the relationship presented in \cite{2009MNRAS.400.2050G}.}
  \label{fig:pismis22_density}
\end{figure}

\cite{Piatti} estimated the distance to Pismis\,\,22 using the E(B$-$V) colour excess and an interstellar absorption value of $A_v \sim 6.0$. Relationships between the hydrogen column density, $N_\text{H}$, and $A_v$ have been previously established. Using the relationship $N_\text{H} = (2.21 \pm 0.09) \times 10^{21} A_v$ from \cite{2009MNRAS.400.2050G}, a column density of $N_\text{H} = 1.3\pm0.2 \times 10^{22}$ cm$^{-2}$ is required to obtain the value of $A_v$ used by \cite{Piatti}. Extracting the average $^{12}$CO(1$-$0) and H\textsc{i} spectra from the extent of Pismis 22 (\citealt{2013A&A...558A..53K}, see Figure\,\,\ref{fig:1614_cs_interest}), we calculate $N_\text{H}$ following \S\ref{sec:line_analysis}. Assuming all the traced gas is at the near distance, we find the total cumulative $N_\text{H}$ as a function of $v_{\text{LSR}}$, which we plot in Figure \ref{fig:pismis22_density}.

From Figure \ref{fig:pismis22_density}, we see that the required value of $N_\text{H}$ is achieved at $v_\text{LSR} \sim -45$ km\,s$^{-1}$, which is similar to the velocity at which the dense open-ring of gas is seen ($\sim -47$ to $-44$ km\,s$^{-1}$). We note that gas traced by CS only appears in a very narrow velocity range, and does not significantly alter the cumulative column density in this region. In addition, we note the conspicuous spatial coincidence between Pismis 22 and the gas ring. Hence, it is possible that the distance to Pismis 22 is currently underestimated, and the stellar cluster is at the distance of this ring seen in component 1 ($\sim 3$ kpc). In this scenario, the dense gas ring may have been the result of the stellar winds from O or B type stars in the cluster, as mentioned previously. A fraction of the energy in these stellar winds could then be accelerating CRs which interact with the nearby ISM to produce the TeV $\gamma$-rays \citep{1982ApJ...253..188V}.

Both the scenarios presented above, namely an undetected SNR and stellar winds from Pismis 22, are consistent with the dense ring traced in CS. The CS ring is likely evidence for stellar winds, which by itself may be behind the TeV emission from HESS\,J1614$-$518 \citep{2008AIPC.1085..241R}. Such stellar winds can be indicative of eventual supernova events, which lends support to a possible undetected SNR acting as a CR accelerator. In these cases, the production of the TeV $\gamma$-ray emission from HESS\,J1614$-$518 is via hadronic interactions of accelerated CRs and the ISM.

\subsection{HESS\,J1616$-$508}
\label{sec:discussion_1616}
Several potential CR accelerators lie near HESS\,J1616$-$508, including SNR Kes 32, SNR RCW 103, and PSR J1617$-$5055. In the following, we discuss the possible relation these candidates have with the observed $\gamma$-ray flux in the context of our analysis of the nearby interstellar medium.

\subsubsection{Kes 32}
Kes 32 (SNR G332.4+00.1), located $\sim17$ arcmin from the centre of HESS\,J1616$-$508, has somewhat uncertain distance and age associations. \cite{2004ApJ...604..693V} used \textit{Chandra} data to study the SNR in X$-$rays, and a single-temperature nonequilibrium ionisation model was used to fit the weak X-ray source spectrum. This was done separately with two background subtractions methods (denoted method 1 and method 2). The results of both methods required a large interstellar absorption column. Method 1 required a hydrogen column density $N_{\text{H}} = 5.6 \pm 0.8 \times 10^{22}$ cm$^{-2}$, while method 2 required column density $N_{\text{H}} = 3.1 \pm 0.4 \times 10^{22}$ cm$^{-2}$. These large column densities suggested that it would be reasonable to associate Kes 32 with the Norma spiral arm. Additionally, OH absorption at $-$88 km\,s$^{-1}$ towards the SNR indicate a distance of at least 6.6 kpc \citep{1975MNRAS.173..649C}. However, this distance estimate is reliant on the Galactic rotation model applied, and using the model presented by \cite{Brand1993} consistent in this study gives a distance of 5.3 kpc.

\begin{figure}[!ht]
\centering
\includegraphics[width=1.0\linewidth]{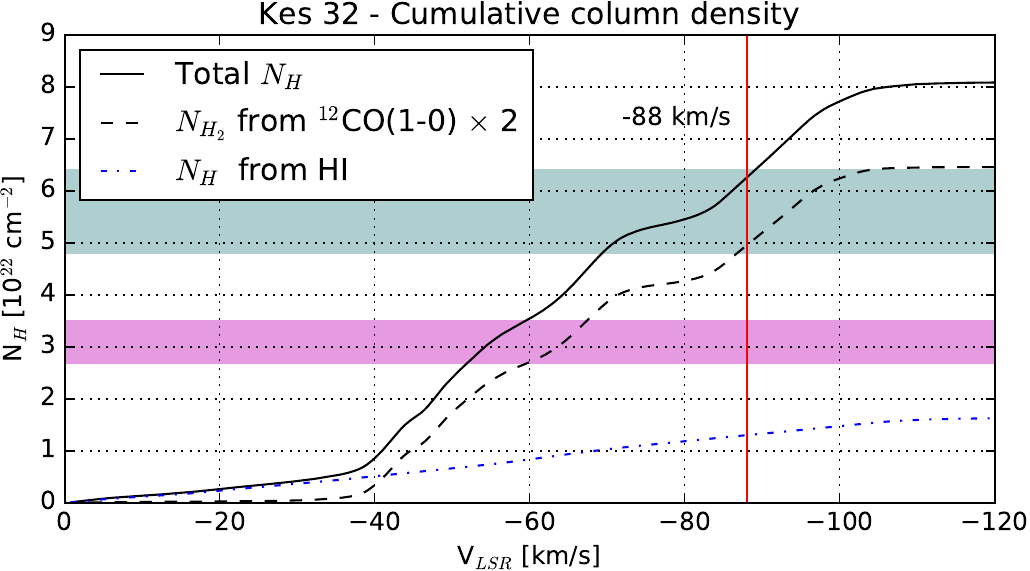}
\caption[what]{Total cumulative hydrogen column density $N_{\text{H}}$ as a function of $v_{\text{LSR}}$ (solid lines) towards the SNR Kes 32 (G332.4+00.1). The cumulative molecular and atomic hydrogen column densities, calculated from $^{12}$CO(1$-$0) and H\textsc{i} data, are shown as dashed and dot-dashed lines respectively. The shaded regions indicate the $N_{\text{H}}$ used to model X-ray spectrum in \cite{2004ApJ...604..693V} (cyan and pink for methods 1 and 2 respectively). Vertical red line marks $v_{\text{LSR}} = -88$ km\,s$^{-1}$.}
  \label{fig:kes32_density}
\end{figure}

We extract the average $^{12}$CO(1$-$0) and H\textsc{i} spectra from within the extent of Kes 32 (as given in \citealt{2014BASI...42...47G}), and use it to calculate $N_{\text{H}}$ following \S\ref{sec:line_analysis}. Assuming that the gas traced is all at the near distance, we find the total cumulative $N_{\text{H}}$ as a function of $v_{\text{LSR}}$. This is displayed in Figure \ref{fig:kes32_density} as a solid black line. The $N_{\text{H}}$ required in methods 1 and 2 presented in \cite{2004ApJ...604..693V} are indicated by the cyan and pink shaded regions respectively. The vertical red line marks $v_{\text{LSR}} = -88$ km\,s$^{-1}$, at which OH absorption is seen \citep{1975MNRAS.173..649C}.

From Figure \ref{fig:kes32_density}, we can see that at $v_{\text{LSR}} = -88$ km\,s$^{-1}$ the total $N_{\text{H}}$ is at the upper limit of the required value in method 1. Thus we suggest that Kes 32 is located at this kinematic velocity $\sim-88$ km\,s$^{-1}$, which would satisfy both the $N_{\text{H}}$ requirement and the OH absorption feature. This velocity would then imply a distance to the SNR of $\sim$ 5.3 kpc following \cite{Brand1993}. This is also consistent with the H\textsc{i} absorption feature coincident with Kes 32 found in the SGPS data as mentioned in \S\ref{sec:results_HI}.

If we assume that Kes 32 is located at this distance, CRs accelerated by the SNR may be interacting with the gas traced in component 5 ($-110$ to $-76$ km\,s$^{-1}$). In a naive scenario in which all the gas traced in component 5 is considered as target material for CRs, the required CR enhancement value $k_{\text{CR}}$ is $\sim 200$ for the observed $\gamma$-ray flux (see Table\,\,\ref{table:Kcr}).

CRs accelerated by Kes 32 may be diffusively reaching the gas overlapping HESS\,J1616$-$508. The centre of Kes 32 is located $\sim17$ arcmin from the centre of HESS\,J1616$-$508, equivalent to $\sim25$ pc at the assumed distance of 5.3 kpc. Figure 1a of \cite{1996A&A...309..917A} displays the modelled values of $k_{\text{CR}}$ at a distance of 30 pc from an impulsive accelerator at a series of time epochs. A $k_{\text{CR}}$ value of $\sim 200$ is possible given a source age of $\sim 10^{3}$ yr, similar to the estimated age of Kes\,\,32 of $\sim3000$ years \citep{2004ApJ...604..693V}. This assumes the diffusion coefficient at an energy of 10 GeV is $D_{10} = 10^{28}$ cm$^{2}$\,s$^{-1}$, which corresponds to relatively fast diffusion. Thus the required $k_{\text{CR}}$ resulting from the diffusion of CRs from Kes 32 is possible, given favourable conditions. However, this scenario assumes that all the gas traced in component 5 is physically connected and is acting as target material. As mentioned in \S\ref{sec:result_1616}, the emission in component 5 is very broad ($\sim 35$ km\,s$^{-1}$) and possibly arises from unresolved overlapping gas features associated with the tangent of the Norma spiral arm. It is then unlikely that all the gas in this component would be acting as CR target material. Additionally there is no striking morphological correspondence between the potential target material and the TeV emission, making it difficult to explain the geometry of the $\gamma$-ray emission based on a hadronic scenario powered by Kes 32.

In a leptonic scenario, CR electrons diffusing from Kes 32 towards HESS\,J1616$-$508 may produce the observed TeV $\gamma$-rays. Assuming a separation of 25 pc at the assumed distance of 5.3 kpc, and using the value of $\overline{n}$ from Table\,\,\ref{table:Kcr}, $\tau_{\text{diff}}$ is calculated to be $\sim 22$ kyr, while $\tau_{\text{sync}} \sim 8$ kyr. As mentioned previously, however, since the gas in component 5 is likely to be distributed along the Norma arm tangent, the value of $\overline{n}$, and consequently the estimated magnetic field strength should be taken as upper limits. If we consider the lower limit of the magnetic field strength within interstellar clouds from \cite{2010ApJ...725..466C} (10 $\mu$G for $\overline{n} \leq 300$ cm$^{-3}$), then $\tau_{\text{diff}}$ and $\tau_{\text{sync}}$ are calculated to be $\sim 16$ kyr and $\sim 25$ kyr respectively. In both cases, $\tau_{\text{diff}}$ is much greater than the estimated age of Kes 32 ($\sim$\,3 kyr) and thus it is unlikely that accelerated electrons would have diffused far enough from the SNR to contribute to the TeV flux of HESS\,J1616$-$508.

\subsubsection{RCW 103}
RCW 103 (SNR G332.4-00.4), located $\sim 13$ arcmin from the centre of HESS\,J1616$-$508, has been well studied in literature. The young SNR is bright in non-thermal X-rays \citep{2015ApJ...810..113F} with an estimated age of $\sim 2000$ years \citep{1984ApJ...284..612N,1997PASP..109..990C}. RCW 103 is located at a distance of $\sim 3.3$ kpc \citep{1975A&A....45..239C,2006PASA...23...69P}, with systematic velocity $\sim -48$ km\,s$^{-1}$ \citep{2006PASA...23...69P}, which places it within the velocity interval of component 2 ($-55$ to $-47$ km\,s$^{-1}$). 

CRs accelerated by this young SNR may be interacting with the gas traced in component 2 to produce TeV $\gamma$-rays. If this is responsible for the $\gamma$-ray flux of HESS\,J1616$-$508, the required $k_{\text{CR}}$ value is $\sim 550$ (Table \ref{table:Kcr}). However, we note that a significant amount of gas, as traced by CO(1$-$0) and CS(1$-$0) emission, cuts across the TeV source as shown in Figures \ref{fig:1616_co} and \ref{fig:1616_cs}. The estimated mass of the gas as traced in CO(1$-$0) within the `bar' is about twice that of the gas within the adopted extent of HESS\,J1616$-$508 (see Tables \ref{table:1616_co_param} and \ref{table:1616_loop_param}). As $k_{\text{CR}}$ is inversely proportional to the mass of CR-target material, if the gas within this bar was acting as target material for CR interaction, the required $k_{\text{CR}}$ value may be as low as $\sim250$.

If the distance to RCW 103 is 3.3 kpc, the separation between the SNR and the centre of HESS\,J1616$-$508 is $\sim 12$ pc. The top panels of Figure 1 in \cite{1996A&A...309..917A} show the $k_{\text{CR}}$ at a distance of 10 pc from an impulsive accelerator, assuming diffusion coefficients of $D_{10} = 10^{26}$ and $10^{28}$ cm$^{2}$\,s$^{-1}$. According to the figures, at a source age of $\sim 10^{3}$ years, similar to the age of RCW\,103 ($\sim 2000$ years), a $k_{\text{CR}}$ of $\sim 300$ is achievable for both $D_{10}$ values.  This implies that the required $k_{\text{CR}}$ may be attained from the diffusion of CRs accelerated by RCW\,103, assuming the gas bar towards HESS\,J1616$-$508 in component 2 is acting as target material.

On the other hand, it is difficult to reconcile the differences in morphology between the target material gas traced in component 2 and the TeV $\gamma$-ray emission of HESS\,J1616$-$508. The bar of gas that cuts across HESS\,J1616$-$508 bears little resemblance to the roughly circular morphology of the TeV $\gamma$-rays. One possible explanation is that accelerated CRs from RCW\,\,103 are interacting with the central region of the dense gas bar towards the middle of the TeV emission. This could be caused by the anisotropic diffusion of CRs preferentially propagating along magnetic field lines \citep{2013MNRAS.429.1643N,2013ApJ...768...73M}. Even so, the differences in gas and $\gamma$-ray morphology make an association between HESS\,J1616$-$508 and RCW 103 somewhat contrived. 

For a leptonic scenario, the diffusion time-scale for CR electrons, $\tau_{\text{diff}}$, is $\sim 4.5$ kyr, based on a separation of $\sim12$ pc between RCW 103 and the centre of HESS\,J1616$-$508, and using the value of $\overline{n}$ from Table\,\,\ref{table:Kcr}. The synchrotron cooling time $\tau_{\text{sync}}$ is $\sim 14$ kyr. While the SNR is estimated to have a young age of $\sim$ 2000 yr, this is not significantly different from the value of $\tau_{\text{diff}}$ given the considerable uncertainties on the value, and it may be possible for CR electrons to have diffusively reached HESS\,J1616$-$508.

\subsubsection{PSR J1617$-$5055}
An association between with HESS\,J1616$-$508, PSR J1617$-$5055 and its corresponding PWN has previously been suggested \citep{Landi2007,Aharonian2008,2011ICRC....6..202T,2013ApJ...773...77A}. Radio dispersion measurements suggest the pulsar is located at a distance between $\sim 6.1 - 6.9$ kpc \citep{Kaspi1998}, which would correspond to a kinematic velocity of $\sim -102$ to $-112$ km\,s$^{-1}$ following \cite{Brand1993}. The assumed distance to the pulsar places it in the velocity range of component 5 ($-110$ to $-76$ km\,s$^{-1}$). PSR J1617$-$5055 is offset from the centre of HESS\,J1616$-$508 TeV emission by $\sim 9$ arcmin, corresponding to $\sim 17$ pc at its assumed distance.

We find no obvious structures in the gas morphology in component 5 that would aid in a pulsar/PWN driven scenario. In the case of a PWN driven TeV source, the ISM is typically seen adjacent to the source, with gas being anti-correlated with the TeV emission (see e.g. \citealt{2001ApJ...563..806B}, \citealt{2016MNRAS.458.2813V}). In our case however, this sort of distribution is not seen in our gas analysis. 

Additionally, the lack of a bow-shock or any asymmetry in X-ray observations of the PWN \citep{Kargaltsev2008} disfavour other scenarios such as a rapidly moving pulsar with high kick velocity \citep{2005Ap&SS.297...93R}. Consequently, there is no convincing evidence to suggest a link between the TeV source with PSR J1617$-$5055.

We note that two other pulsars are seen towards HESS\,J1616$-$508; PSR\,J1616$-$5109 and PSR\,J1614$-$5048. However, neither of these pulsars have been considered likely counterparts in previous studies, stemming from their modest spin down power and relatively large offsets from the TeV source \citep{Landi2007,2012ApJ...756....5L,2017ApJ...841...81H}.

\subsubsection{An accelerator at the centre of HESS\,J1616$-$508?}
Based on our ISM studies, the known accelerators in the nearby regions towards HESS\,J1616$-$508 have some issues in explaining the observed TeV emission. As mentioned in \S\ref{sec:results}, the gas in component 1 ($-47$ to $-39$ km\,s$^{-1}$) towards HESS\,J1616$-$508 traced in $^{12}$CO(1$-$0) emission forms a molecular cloud structure that appears to overlap the TeV source (Figure \ref{fig:1616_co}). Additionally, there appears to be a circular void-like feature towards the centre of HESS\,J1616$-$508. This void is also present in HEAT [CI] data (Figure \ref{fig:HEAT}), as well as being very pronounced in the $^{13}$CO(1$-$0) emission.

We now postulate on a previously undetected accelerator at the centre of HESS\,J1616$-$508, interacting with this conspicuous gas traced in component 1. The void may be associated with some as-of-yet undetected accelerator which has blown out a cavity in the gas. This accelerator may then be the source of high energy CRs responsible for $\gamma$-ray emission from HESS\,J1616$-$508. Looking at the integrated $^{12}$CO(1$-$0) and $^{13}$CO(1$-$0) images (Figure \ref{fig:1616_co}), the diameter of the void is $\sim 0.1$ to 0.2 degrees. At a kinematic distance of $\sim 3$ kpc, would correspond to $\sim 5$ to 10 pc. According to Table\,\,\ref{table:Kcr}, the required CR enhancement factor for the gas in component 1 is of the order $\sim 300$. A young impulsive accelerator, such as a SNR, located coincident with the gas void would be readily able to supply the required $k_{\text{CR}}$ value \citep{1996A&A...309..917A}. In this scenario, the void in the ISM may have been blown out by a progenitor star.

We also note the peculiar line of gas in component 1 which points towards the peak of the TeV emission. This line can be seen in $^{12}$CO(1$-$0) (Figure \ref{fig:1616_co}) and [CI] (Figure \ref{fig:HEAT}), but appears most prominently in $^{13}$CO(1$-$0).

A similar thin line of molecular gas has been observed pointing towards another Galactic TeV source, HESS\,J1023$-$575 \citep{2009PASJ...61L..23F}. The formation of this molecular feature has been speculated to be caused by an energetic event such as an anisotropic supernova explosion. It may be the case for HESS\,J1616$-$508 that the thin line of gas in component\,1 has been formed under similar circumstances. Molecular jets have been seen towards binary systems, such as the microquasar SS\,433 \citep{2008PASJ...60..715Y}, which are thought to be accretion powered. However, in our case of HESS\,J1616$-$508, no suitable counterpart has been detected for such a scenario.

\section{Conclusions}
Using 3\,mm data from the Mopra Radio telescope, 7\,mm data from Mopra and the ATCA, archival H\textsc{i} data, as well as [CI] data from HEAT, we have studied the interstellar medium towards two unidentified TeV sources, HESS\,J1614$-$518 and HESS\,J1616$-$508.

Towards HESS\,J1614$-$518, CO(1$-$0) observations from Mopra reveal diffuse molecular gas at several velocities along the line-of-sight that appear to overlap the TeV source. While the morphological correspondence with the TeV emission is not particularly strong, $^{12}$CO(1$-$0) in component 1 ($-50$ to $-40$ km\,s$^{-1}$) is seen to overlap most of the TeV emission. 7\,mm observations from Mopra in the CS(1$-$0) tracer revealed a peculiar open ring-like structure of dense gas towards the centre of HESS\,J1614$-$518, located at a velocity consistent with the $^{12}$CO(1$-$0) emission seen in component 1, and was the only dense gas feature seen overlapping the TeV source.

CO(1$-$0) observations towards HESS\,J1616$-$508 revealed multiple components of diffuse molecular gas overlapping the TeV source. Particularly interesting features included a loop of gas cutting through the TeV source in component 2 ($-55$ to $-47$ km\,s$^{-1}$), and positionally coincident molecular gas overlapping the entire TeV source in component 1 ($-47$ to $-39$ km\,s$^{-1}$). Additionally, in component 1, there appeared to be a void in the diffuse gas towards the central TeV peak. Dense gas was traced in CS(1$-$0) with a similar morphology and velocity as the loop feature seen in the diffuse gas in component 2.

We estimated the physical parameters of the gas using the CO, CS and H\textsc{i} data for the gas components and interesting features seen towards HESS\,J1614$-$518 and HESS\,J1616$-$508. For hadronic scenarios, assuming the gas in the diffuse components were acting as target material, it was found that the required values for the total CR energy budget $W_p$ were $\sim10^{48}$ erg and $\sim10^{47}$ to $10^{48}$\,erg for HESS\,J1614$-$518 and HESS\,J1616$-$508 respectively. The required CR enhancement factors were calculated based on total gas masses for each of the gas components, and are displayed in Table \ref{table:Kcr}.

For HESS\,J1614$-$518, we find that the scenario involving an as-of-yet undetected SNR, potentially associated with the X-ray sources Suzaku Src A and XMM-Newton Src B1, could generate the observed TeV $\gamma$-rays in a hadronic interaction scenario. 

The stellar wind scenario involving the stellar cluster Pismis 22 at the estimated distance of $\sim 1$ to 2 kpc was more difficult to reconcile, given the lack of gas seen at the corresponding $v_\text{LSR}$. However, the total column density towards the cluster and the spatial coincidence with the dense gas ring seen in CS(1$-$0) at $\sim 3$ kpc suggests that the distance to Pismis may be underestimated. A stellar wind scenario driven by O and B type stars in the cluster and interacting with the gas traced in component\,1 may then be contributing to the observed TeV $\gamma$-ray flux.

Several accelerator candidates towards HESS\,J1616$-$508 were investigated in light of our ISM study. Neither of the two young SNRs that flank the TeV source, Kes 32 and RCW 103, were found to be strong candidates for association. We also found no convincing evidence to link PSR J1617$-$5055 and its associated PWN to TeV $\gamma$-rays from HESS\,J1616$-$508. Due to the somewhat conspicuous nature of the diffuse gas seen in  component 1 ($-47$ to $-39$ km\,s$^{-1}$), we speculate on an undetected accelerator at the centre of the TeV source interacting with said gas. We find that a CR accelerator such as a young SNR would readily be able explain the TeV $\gamma$-ray flux from HESS\,J1616$-$508.

Based on our study of the ISM, we find no conclusive evidence to link either HESS\,J1614$-$518 or HESS\,J1616$-$508 to any known counterparts. However, the angular resolutions of next-generation $\gamma$-ray telescopes, such as the Cherenkov Telescope Array, will approach that of this ISM study. This would enable more detailed morphological comparisons in the future between the TeV $\gamma$-ray emission and the interstellar gas, allowing for a better understanding of these two mysterious sources. 

\begin{acknowledgements}
The Mopra radio telescope is part of the Australia Telescope National Facility. Operations support was provided by the University of New South Wales and the University of Adelaide. The UNSW Digital Filter Bank used for the observations with Mopra was provided with financial support from the Australian Research Council (ARC), UNSW, Sydney and Monash universities. We also acknowledge ARC support through grants DP120101585 and LE160100094. J.C.L. and S.P. acknowledge support through the provision of Australian Government Research Training Program Scholarships. The HEAT telescope is financially supported by the National Science Foundation under award numbers ANT-0944335 and AST-1410896, with additional funding from the Australian Government’s Australian Antarctic Science Grant Program and NCRIS, and with logistics through the United States Antarctic Program.
\end{acknowledgements}

\bibliographystyle{pasa-mnras}
\bibliography{references.bib}

\begin{appendix}

\begin{figure*}[!ht]
\centering
\includegraphics[width=\linewidth]{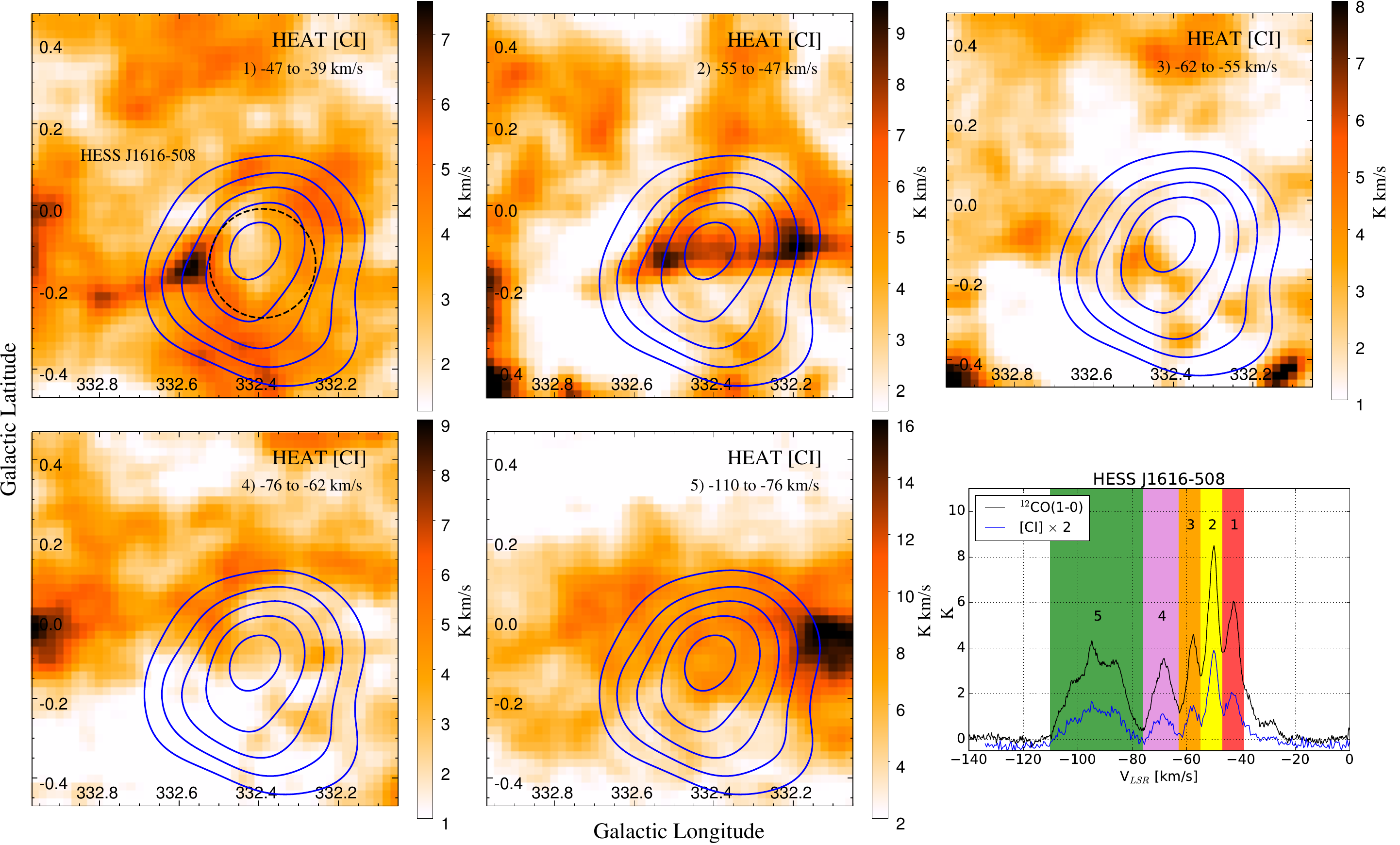}
\caption[what]{HEAT CI (J=2$-$1) integrated intensity images [K\,km\,s$^{-1}$] within the labelled velocity intervals towards HESS\,J1616$-$508. Overlaid are HESS excess counts contours (blue) at the 30, 45, 60, 75 and 90 levels. The dashed black circle in the top left panel is the RMS extent of HESS\,J1616$-$508 \citep{hess_plane}. The bottom-right panel shows average spectra of $^{12}$CO(1$-$0) (black) and [CI] (blue) emission within the circular region. For clarity, the [CI] spectrum is scaled by a factor of 2. The velocity intervals used in the integrated image panels are indicated by the shaded rectangles.}
  \label{fig:HEAT}
\end{figure*}

\begin{figure*}[!ht]
\centering
\includegraphics[width=\linewidth]{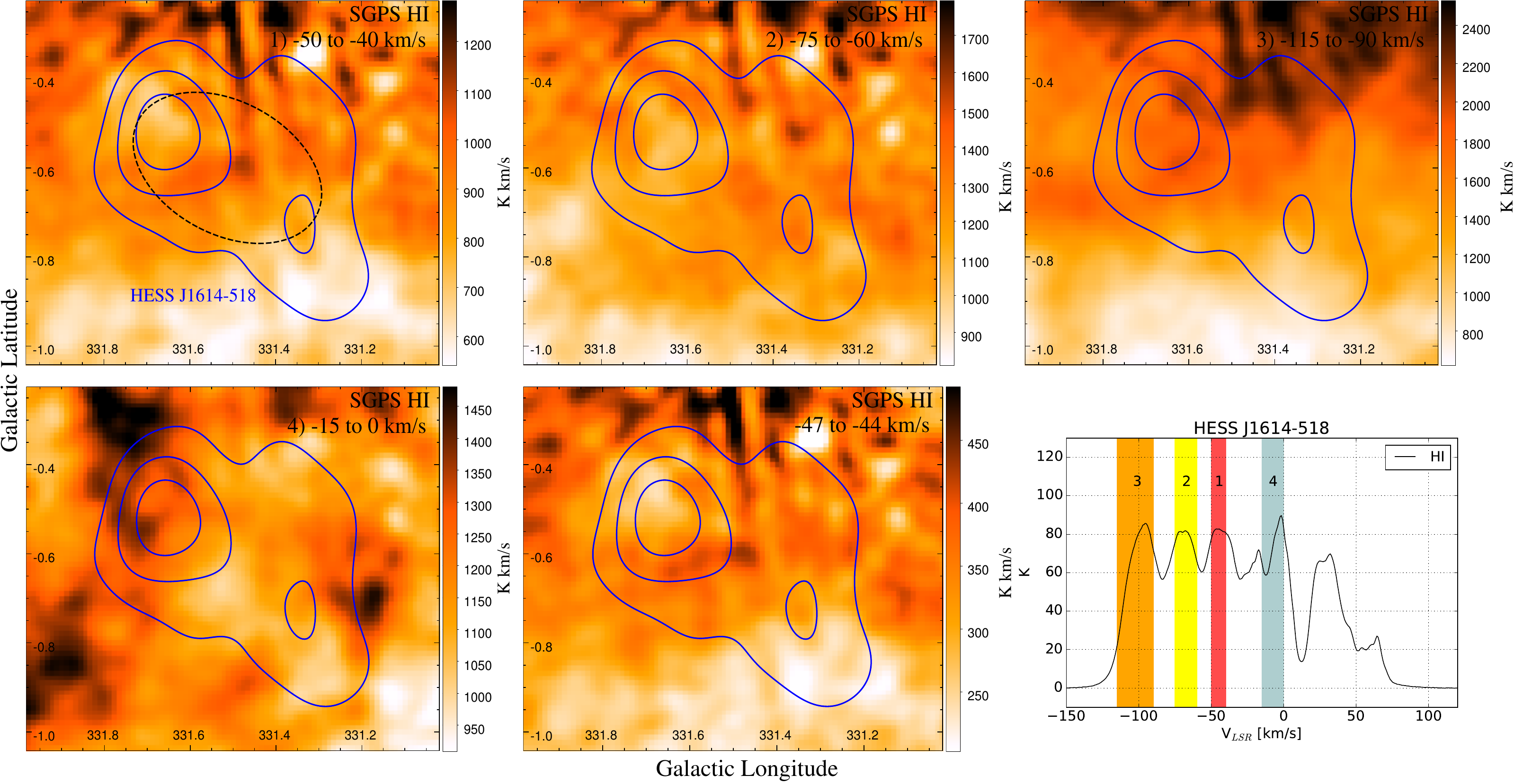}
\caption[what]{H\textsc{i} integrated intensity images [K\,km\,s$^{-1}$] from SGPS data within the labelled velocity intervals towards HESS\,J1614$-$518. Overlaid are HESS excess counts contours (blue) at the 30, 45 and 60 levels. The dashed black ellipse is the RMS extent of HESS\,J1614$-$518 \citep{hess_plane}. The bottom-right panel which shows the average H\textsc{i} emission spectrum within the extent of the TeV source. The velocity integration intervals for each component as described in text (\S\ref{sec:result_1614}) is indicated by the shaded rectangles.}
  \label{fig:SGPS_1614}
\end{figure*}

\begin{figure*}[!ht]
\centering
\includegraphics[width=\linewidth]{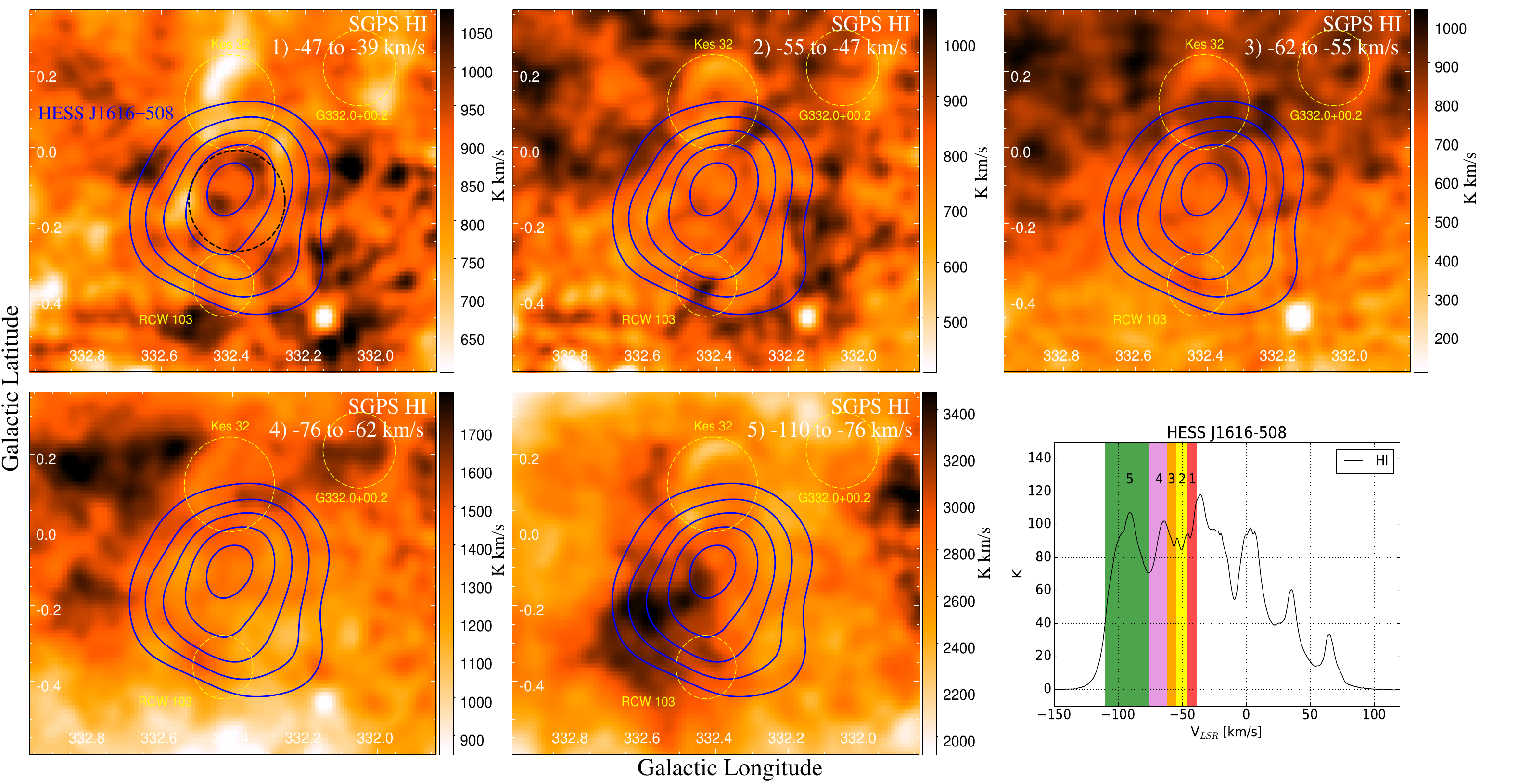}
\caption[what]{H\textsc{i} integrated intensity images [K\,km\,s$^{-1}$] from SGPS data within the labelled velocity intervals towards HESS\,J1616$-$508. Overlaid are HESS excess counts contours (blue) at the 30, 45, 60, 75 and 90 levels. The dashed black circle is the RMS extent of HESS\,J1616$-$508 \citep{hess_plane}. The yellow dashed circles indicates the positions of known SNRs in the region \citep{2014BASI...42...47G}. The bottom-right panel shows the average H\textsc{i} emission spectrum within the extent of the TeV source. The velocity integration intervals for each component as described in text (\S\ref{sec:result_1616}) is indicated by the shaded rectangles.}
  \label{fig:SGPS_1616}
\end{figure*}

\end{appendix}

\end{document}